\documentclass[twocolumn]{aastex631}

\usepackage{graphicx}

\usepackage{xcolor}
\usepackage{longtable}
\usepackage{subfigure}
\usepackage{tikz}
\usepackage{csquotes}
\usepackage{lineno}

\usepackage{tabularx}
\usepackage{amssymb}
\usepackage{orcidlink}

\usetikzlibrary{decorations.pathreplacing,positioning, arrows.meta}

\newcommand{\Qgamma}{Q_{\gamma}}

\begin{document}

\title{Deep Multimessenger Search for Compact Binary Mergers in LIGO, Virgo, and Fermi/GBM Data from 2016-2017}

\author{Marion Pillas \orcidlink{0000-0003-3224-2146}}
\affiliation{Universit\'e Paris-Saclay, CNRS/IN2P3, IJCLab, F-91405 Orsay, France}
\email{marion.pillas@ligo.org}

\author{Tito \surname{Dal Canton} \orcidlink{0000-0001-5078-9044}}
\affiliation{Universit\'e Paris-Saclay, CNRS/IN2P3, IJCLab, F-91405 Orsay, France}

\author{Cosmin Stachie}
\affiliation{Universit\'e C\^ote d’Azur, Observatoire de la C\^ote d’Azur, CNRS, Artemis, F-06304 Nice, France}

\author{Brandon Piotrzkowski \orcidlink{0000-0001-8919-0899}}
\affiliation{University of Wisconsin-Milwaukee, Milwaukee, WI 53201, USA}

\author{Fergus Hayes \orcidlink{0000-0001-7628-3826}}
\affiliation{SUPA, School of Physics and Astronomy, University of Glasgow, Glasgow G12 8QQ, UK}

\author{Rachel Hamburg \orcidlink{0000-0003-0761-6388}}
\affiliation{Universit\'e Paris-Saclay, CNRS/IN2P3, IJCLab, 91405 Orsay, France}

\author{Eric Burns \orcidlink{0000-0002-2942-3379}}
\affiliation{Department of Physics \& Astronomy, Louisiana State University, Baton Rouge, LA 70803, USA}

\author{Joshua Wood \orcidlink{0000-0001-9012-2463}}
\affiliation{NASA Marshall Space Flight Center, Huntsville, AL 35805, USA}

\author{Pierre-Alexandre Duverne \orcidlink{0000-0002-3906-0997}}
\affiliation{Universit\'e Paris-Saclay, CNRS/IN2P3, IJCLab, 91405 Orsay, France}
\affiliation{Universit\'e Paris Cit\'e, CNRS, Astroparticule et Cosmologie, F-75013 Paris, France}

\author{Nelson Christensen \orcidlink{0000-0002-6870-4202}}
\affiliation{Universit\'e C\^ote d’Azur, Observatoire de la C\^ote d’Azur, CNRS, Artemis, F-06304 Nice, France}

\begin{abstract}
GW170817-GRB 170817A provided the first observation of gravitational waves from a neutron star merger with associated transient counterparts across the entire electromagnetic spectrum. This discovery demonstrated the long-hypothesized association between short gamma-ray bursts and neutron star mergers. More joint detections are needed to explore the relation between the parameters inferred from the gravitational wave and the properties of the gamma-ray burst signal. We developed a joint multimessenger analysis of LIGO, Virgo, and Fermi/GBM data designed for detecting weak gravitational-wave transients associated with weak gamma-ray bursts. As such, it does not start from confident (GWTC-1) events only. Instead, we take the full list of existing compact binary coalescence triggers generated with the PyCBC pipeline from the second Gravitational-Wave Observing Run (O2), and reanalyze the entire set of public Fermi/GBM data covering this observing run to generate a corresponding set of gamma-ray burst candidate triggers. We then search for coincidences between the gravitational-wave and gamma-ray burst triggers without requiring a confident detection in any channel. The candidate coincidences are ranked according to a statistic combining each candidate's strength in gravitational-wave and gamma-ray data, their time proximity, and the overlap of their sky localization. The ranking is then converted to a false alarm rate using time shifts between the gravitational-wave and gamma-ray burst triggers. We present the results using O2 triggers, which allowed us to check the validity of our method against GW170817-GRB 170817A. We also discuss the different configurations tested to maximize the significance of the joint detection.
\end{abstract}


\section{Introduction}

Since 2015, the year of the first gravitational-wave (GW) detection GW150914 \citep{GW150914}, the Advanced LIGO \citep{LIGO} and Advanced Virgo \citep{Virgo}, L-shaped kilometer-scale Michelson laser interferometers, have continued their searches to detect GW events. Sources such as compact binary coalescences (CBCs) are candidates for detection. Possible counterparts of CBCs involving at least one neutron star, for example binary neutron stars (BNSs) and neutron star-black hole (NSBH) mergers, are short gamma-ray bursts (sGRBs) \citep{Eichler_1989, Nakar_2007}. GRBs are bursts of highly energetic gamma rays with a duration ranging from less than a second to several minutes. When a BNS or an NSBH merger occurs, an accretion disk may be formed around the resulting black hole and an sGRB jet can be ejected. The fraction of BNS and NSBH mergers leading to GRB emission is currently not precisely known. GRBs are usually classified into two categories based on their duration, represented by the $T_{90}$ statistic, which is instrument-dependent and defined as the time interval in which the integrated photon count increases from 5\% to 95\% of the total counts above the background. The spectral hardness of the prompt emission also contributes to the classification: long GRBs tend to be soft while short GRBs are hard~\citep{kouveliotou_1993}. Long GRBs, having a duration $T_{90} \geq 2$ s, are typically associated with a subclass of core-collapse supernovae (CCSNe) \citep{Woosley_1993,Cano_2017}. sGRBs are believed instead to originate mostly in CBC systems containing at least one neutron star, as demonstrated by the joint detection GW170817-GRB 170817A \citep{GW170817, Abbott_2017, LIGOScientific:2017zic}.
There are, however, important exceptions to this classification \citep{Ahumada2021, Rastinejad2022}.

To answer fundamental questions about sources of GWs and GRBs, for instance to determine the fraction of sGRBs associated with neutron star mergers, or the mechanisms responsible for the formation of a jet, a significantly larger sample of joint GW/GRB observations is needed.
To this end, there are currently several independent approaches to searching for joint GW/GRB associations, such as PyGRB (\citep{PhysRevD.83.084002, joint-searches, Williamson_2014}) and X-pipeline (\citep{X-pipeline, PhysRevD.86.022003, joint-searches}) which assume that there is a GRB detection and perform a deep search for a nearby GW signal (or more generally GW transients in the case of X-pipeline). On the low-latency time scale, RAVEN (\citep{RAVEN-Urban, joint-searches, 2022APS..APRY13002P}) assumes a GW and an electromagnetic (EM) detection and checks for their compatibility in time and sky location. 
By contrast, in the analysis presented in \citep{2022AAS...24034801F}, they start from the GW events from the GWTC-3 catalog and they look for potential associations with Fermi-GBM and Swift-BAT data. 
Other analyses have been performed for specific events, such as Blackburn's method \citep{Blackburn2015} to search for a counterpart to GW150914 \citep{Connaughton:2016umz}. 
The LIGO/Virgo O1 - Fermi/GBM \citep{O1GBM} analysis also computes a false alarm probability (FAP) assuming a GW detection and checks for a nearby GBM signal.
Moreover, a method to search for associations between insignificant GW and GRB candidates during the first GW observing run was proposed in \citep{Nitz_2019} and a potential association, named \textquote{1-OGC 151030,} was found.

Finally, during O2, a joint analysis calculates a \emph{p-value} for the association of GW170817 and GRB 170817A, assuming a GW detection and a GBM detection and checking for compatibility in time and sky localization. 
However, many of these searches have statistical or computational limitations that prevent their application to a large number of weak candidate events.
For example, both X-pipeline and PyGRB require hours to days (running on a CPU cluster) to analyze the GW data around the time of a single GRB and start from the assumption that the GRB is a robust detection.
With RAVEN the joint false alarm rate (FAR [yr$^{-1}$]) of the foreground association is proportional to the single FAR in each channel, leading to statistical limitations in the case of sub-threshold events. However, recent versions of RAVEN can also deal with events of not particularly high confidence \citep{2022APS..APRY13002P} by changing its calculation of the joint FAR. \\

Motivated by the possibility that sGRBs come from CBCs, and by the joint observation of GW170817 \citep{GW170817} and GRB 170817A \citep{GRB170817A_1,GRB170817A_2} coming from a common BNS source, we develop a method to search for coincidences between transient events in the Fermi/GBM data and LIGO O2 CBC triggers, in order to increase the number of joint GW/GRB detections. This is a deep search in the sense that it does not consider confident events only. Instead, for the analysis presented in this paper we use the full list of Fermi/GBM triggers (here 780,206 triggers) and GW triggers (here 15,270) coming from O2. The goal is to find pairs of Fermi/GBM and GW triggers that could possibly have a common origin, rank the pairs with a ranking statistic, and assign a FAR to them. The statistic we use here is similar to the ranking statistic in \citep{Ranking}, which we describe in Section \ref{RankingStat}, but different configurations are tested to compute the terms of this statistic and rank the associations. 

The organization of the paper is as follows. In Section~\ref{method} we describe the method used to search for GW/GBM pairs. In Section~\ref{results} we detail the different configurations tested in this analysis and the results we obtained with each configuration. Section~\ref{conclusion} contains a discussion about the main results of this search, and proposes improvements that we can implement in the method for future searches.

\section{Method}
\label{method}
In the following section, we describe how we rank the potential associations. First the triggers are generated from the GW data and the gamma-ray data independently. Then we find all the possible associations and calculate a ranking statistic based on the properties we have on these triggers (GW and GBM triggers' significance, time and sky proximity between them). Finally, we estimate the statistical significance of each pair. The following subsections present how we generate the GW and gamma-ray candidates and detail the computation of the different terms of the ranking statistic. 

\subsection{Gravitational-wave Candidate Generation}
The search for GW signals produced by a CBC is carried out by several independent pipelines using different techniques to improve detection efficiency and to compare the recorded data with theoretical signals derived from general relativity.
Since the parameters of each GW source are not known in advance and must be inferred from the data, the search explores the CBC source parameter space, usually composed of the masses and spins of the components of the binary, by covering it with a grid of model waveforms called a \emph{template bank} \citep{https://doi.org/10.48550/arxiv.1705.01845}. 
In the following analysis we use O2 triggers from the PyCBC pipeline \citep{PyCBC}, which is a modeled matched-filtering-based \citep{Allen_2012} analysis pipeline that identifies CBC events by correlating data with a template bank of waveforms. There are several steps in the matched-filtering techniques used by the pipeline
to distinguish noise from signals and measure their significance. The data are collected from all the available detectors and are then scanned to find matches with the waveforms in the template bank. A signal-to-noise ratio (S/N or $\rho$) time series is computed to find times when the S/N exceeds a predetermined threshold, a CBC trigger is generated at each maxima in the time series. One major problem that can interfere with GW detection is the instrumental or environmental noise that severely limits the sensitivity of the GW interferometric detectors. This pipeline implements veto techniques to reject these transient non-Gaussian noise \emph{glitches} that create large S/N values but can be easily discriminated from CBC signals. Many other strategies, such as the $\chi^2$ test or the reweighted S/N defined in Equation~6 in \citep{GW150914}, are used to distinguish signals from these glitches. Finally, to compute the significance of the resulting triggers, which are described by a FAR value representing how often we expect noise to produce a trigger with the same ranking statistic as the candidate in question, a background sample is generated using a time-shift method \citep{W_s_2009} between the data of the available detectors. The background generation in our analysis is based on the same time-shift method. 

In this analysis, we focus on O2 PyCBC coincident triggers from the two LIGO detectors (LIGO Hanford (H1) and LIGO Livingston (L1)) in which potential binary black hole, BNS, and NSBH signals can be found. 
For each GW trigger, we compute the posterior distribution for the sky location of the source, under the assumption that the trigger is of astrophysical origin.
We use \texttt{Bayestar} \citep{Bayestar} due to its simplicity and low computational cost, through the \verb|pycbc_make_skymap| tool.
Although the search triggers are produced from LIGO data only, we also use Virgo data (when available) for the sky localization, as they can lead to significantly narrower posterior distributions.

\subsection{Gamma-ray Candidate Generation}
The Fermi Gamma-ray Space Telescope \citep{Michelson_2010} is a space observatory launched on 2008 June 11 and dedicated to the detection of the most energetic phenomena taking place in the Universe through observations of gamma-ray radiation. Aboard Fermi, the Gamma-ray Burst Monitor (GBM; \citep{Meegan_2009}) instrument is used to observe GRBs coming from the entire sky not occulted by the Earth within the energy range of 8 keV--40 MeV. \
The GBM flight hardware comprises 14 NaI(TI) detectors, which are used to measure the low-energy spectrum from 8 keV to 1 MeV, and two bismuth germanate detectors with an energy range of $\sim$200 keV to $\sim$40 MeV. An onboard \textquote{trigger} occurs when there is an increase in the count rates of two or more NaI detectors above an adjustable threshold. The GBM flight software contains algorithms to compute the location of trigger events based on the relative rates in the NaI detectors and to evaluate the probability that a trigger arises from a GRB. In the analysis presented here, we do not use the onboard triggers; instead we use the time-tagged event (TTE) data candidate gamma-ray signals from ground searches of GBM data. The TTE data are time series of photon “counts” for which the time and energy are recorded. 

The GBM trigger generation is realized by the so-called GBM Targeted Search \citep{O1GBM, https://doi.org/10.48550/arxiv.1903.12597}.
This produces triggers by searching for an excess of photon counts compatible with GRBs over a variety of overlapping time windows, covering durations from 64 ms to 8.192 s.
A log-likelihood ratio (LLR) is computed for each time window, allowing for the generation of GBM triggers labeled with a duration, a mission elapsed time (MET), and an LLR.
The LLR calculation uses photon rates produced by a GRB in the GBM detectors that depend on the energy channels in a way that can be predicted after a particular spectral shape has been assumed for the GRB. 
Finally, the GBM Targeted Search also contains a clustering algorithm that only keeps the trigger in regions of the duration-time plane that have the highest LLR if LLR$\geq$5.
This threshold is justified in \citep{Ranking}.

We use the clustered GBM Targeted Search results in the analysis presented here.
This is justified by our desire to remove the correlation between the triggers and also by the high cost of dealing with the unfiltered triggers.
The result of the Targeted Search also gives us the sky localization of the GBM triggers with a sky grid resolution of 5\textdegree.
The search produces maps of the posterior probability distribution $P(\Omega|D_\gamma)$, which quantifies the probability that the GBM trigger comes from a specific location $\Omega$ given the data $D_\gamma$. Note that $P(\Omega|D_\gamma)$ may be nonzero over the portion of sky occulted by the Earth; a large $P(\Omega|D_\gamma)$ over the Earth indicates that the GBM trigger is likely to be terrestrial. 

\begin{figure}
    \centering
     \begin{subfigure}
         \centering
         \includegraphics[width=0.45\textwidth]{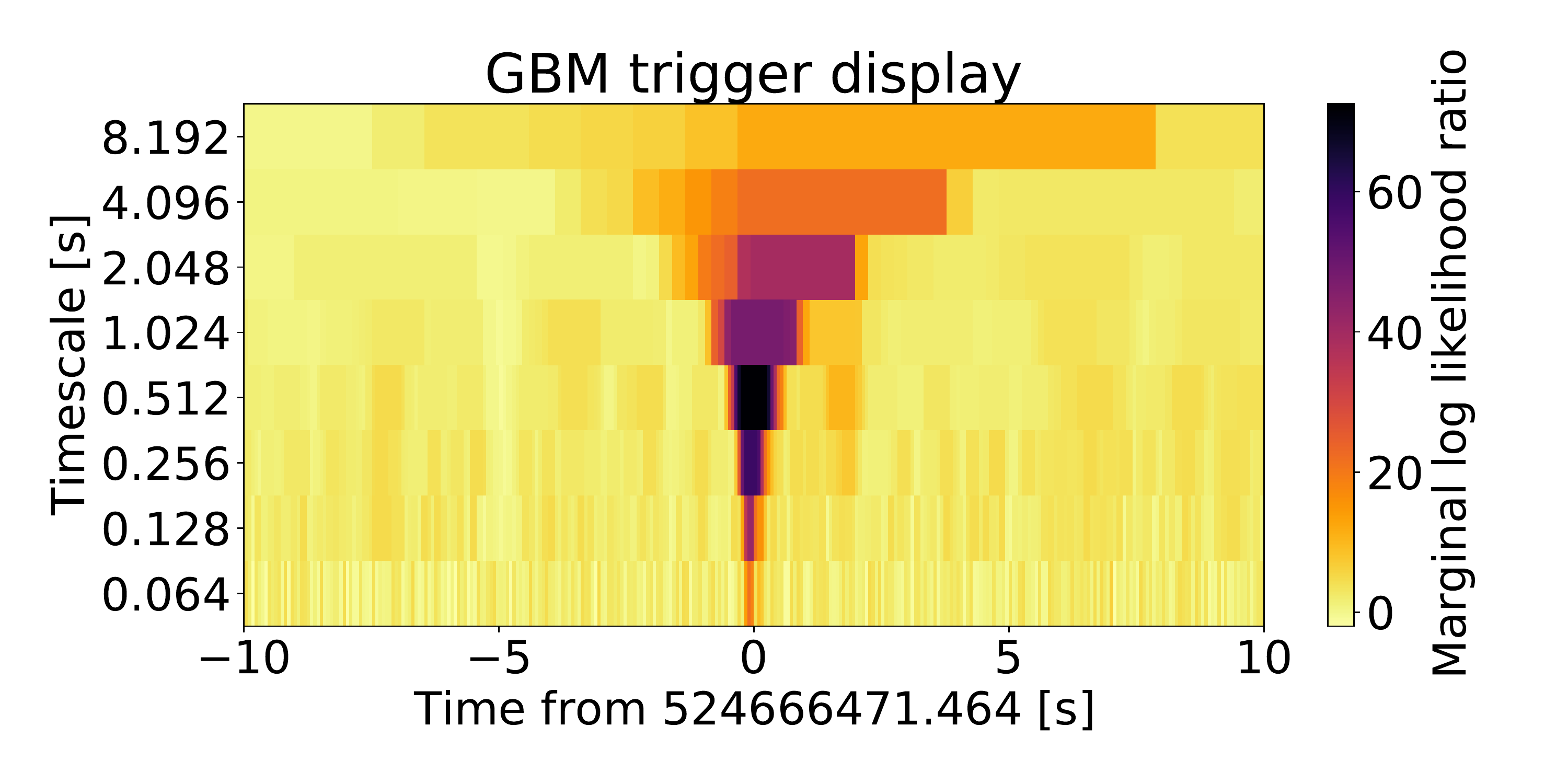}
         \label{fig:waterfallGRB170817A}
     \end{subfigure}
     \hfill
     \begin{subfigure}
        \centering
        \includegraphics[width=0.45\textwidth]{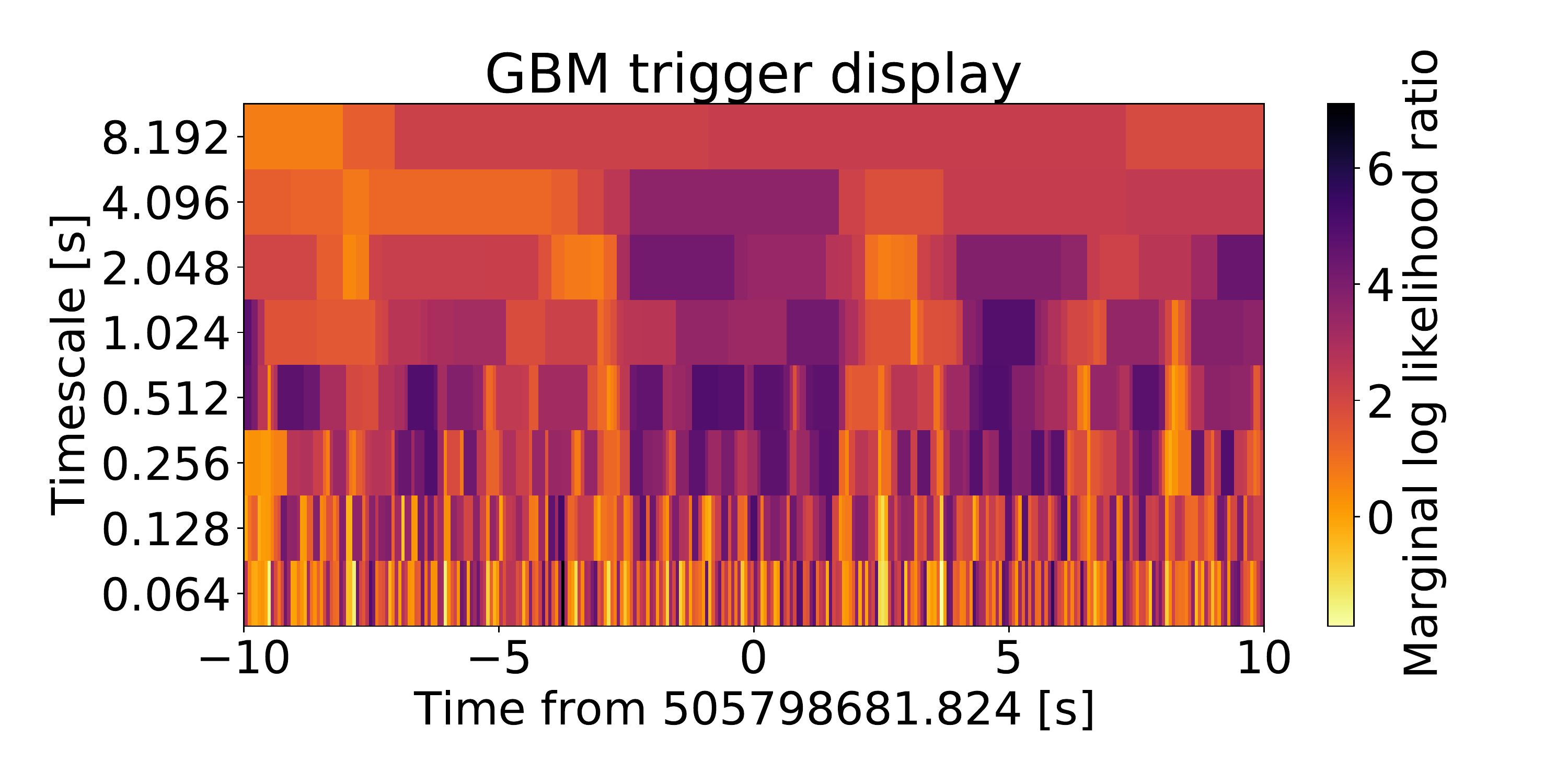}
        \label{fig:waterfallnoise}
     \end{subfigure}
    \caption{Difference in the display of 20 s windows over the search result for GRB 170817A (top) and background only (bottom).}
    \label{fig:waterfall}
\end{figure}

The search results can be displayed with a waterfall plot.
The top panel of Figure~\ref{fig:waterfall} on top shows the display of GRB 170817A.
The x-axis represents the GBM MET and the y-axis shows the different duration windows.
The color of each rectangle denotes the LLR value.
An sGRB shows up as a particular waterfall shape, due to the progressive drop of the LLR as the search window deviates more and more from the best-fitting time and duration.
Far from the sGRB, displayed in the bottom panel of Figure~\ref{fig:waterfall}, one can also see the statistical fluctuations of the persistent, slowly varying and ever-present GBM background.
This background can be divided into a photon component and a charged particle component.
The former is produced by the actual overall photon flux coming from the ensemble of sources in the celestial sphere around Fermi, such as the Sun, the Earth, the cosmic gamma-ray background, and gamma-ray pulsars.
The latter includes cosmic rays and any charged particle interacting with the satellite materials or with the photomultipliers.

\subsection{Ranking Statistic}
\label{RankingStat} 
Addressing joint detections, there are many statistics to rank the potential associations. We first consider two simple ranking statistics, inspired from \citep{Nitz_2019}: 
\begin{itemize}
    \item \emph{Naive time statistic}: a naive product of GW and GBM likelihoods, including the time overlap, and summarized by 
    \begin{equation}
        \mathrm{ln}(\Lambda) = \frac{{\hat{\rho}_{g}}^2}{2} + \mathrm{LLR} + \mathrm{ln}(I_{\Delta \mathrm{t}})
        \label{eqn:naivetime}
    \end{equation}
    \item \emph{Naive time and sky statistic}: a naive product of GW and GBM likelihoods, including the time and skymap overlaps and summarized by
    \begin{equation}
        \mathrm{ln}(\Lambda) = \frac{{\hat{\rho}_{g}}^2}{2} + \mathrm{LLR} + \mathrm{ln}(I_{\Delta \mathrm{t}}) + \mathrm{ln}(I_{\Omega})
        \label{eqn:naivetimesky}
    \end{equation}
\end{itemize}
Here $\hat{\rho}_{g}$ is the network-reweighted S/N \citep{PyCBC} described by Equation~\ref{eqn:coincidentSNR}: 
\begin{equation}
\hat{\rho}_g = \sqrt{\sum_{d}^{}\hat{\rho}_d^2}
\label{eqn:coincidentSNR}
\end{equation}
where $d$ represents the available GW detectors. The term $\frac{\hat{\rho}_g^2}{2}$ represents the log-likelihood ratio $\frac{\mathbb{P}(\mathrm{data}|\mathrm{noise, signal})}{\mathbb{P}(\mathrm{data}|\mathrm{noise})}$ on the GW side, following the computation from \cite{Nitz_2017}. 
$I_{\Delta t}$ and $I_{\Omega}$ are Bayes factors that quantify the overlap of the posterior distributions for the arrival times (time offset) and sky locations (skymap overlap) inferred separately from the GW and GBM data.
These ranking statistics are straightforward to implement, and are presented here in Section~\ref{Preliminary}. We also justify our choice to not use them in this work and to compute the Bayesian ranking statistic described in the following paragraphs.

When searching for signals in GW and GBM data, the two simple possibilities of noise and versus noise+signal become a complicated multimessenger rainbow of mutually exclusive hypotheses:
\begin{itemize}
    \item[1.] ($H_{NN}$) hypothesis: both GW and GBM data contain their respective noise only. 
    \item[2.] ($H_{NS}$) hypothesis: GW data contain noise and signal, and GBM contains only noise.
    \item[3.] ($H_{SN}$) hypothesis: GW data contain noise, and GBM contains noise and signal.
    \item[4.] ($H_{SS}$) hypothesis: both GW and GBM contain noise and signal, but the signals are unrelated.
    \item[5.] ($H_{C}$) hypothesis: both GW and GBM contain noise and signal, and the signals come from the same source.
\end{itemize}
Here by "noise" in the GW data we also consider the possibility of Gaussian noise with a glitch on top. To make a frequentist multimessenger detection in the usual way, we combine cases 1--4 into a null hypothesis and then test hypothesis 5 versus null. As done in many astrophysical searches \citep{Piotrzkowski_2022}, we then define a joint ranking statistic from the GW and GBM data, calculate the corresponding background distribution and finally produce a $p$-value or FAR. Here hypotheses 1 to 4 in the list above can be defined as situations that all represent the \textquote{background}.

When ranking the different GW--GBM candidate combinations, we make the following considerations: 
\begin{itemize}
    \item A larger GW candidate significance corresponds to a more likely candidate.
    \item A larger GBM candidate significance corresponds to a more likely candidate in the GBM channel.
    \item The closer in time the GBM candidate is to the GW merger time, the more likely the candidate is. 
    \item The more the sky localizations overlap, the more confident the candidate is.
\end{itemize}

Going from the list of different hypotheses, and by a Bayesian derivation, we arrive at the association ranking statistic calculated in \citep{Ranking} and derived from \citep{SKYOVERLAP}, which can be summarized with the following formula: 
\begin{equation}
\Lambda =\frac{P(D_{g}, D_{\gamma}|H_{C})}{P(D_{g}, D_{\gamma}|H_{NN} \bigvee H_{SN} \bigvee H_{NS} \bigvee H_{SS})} ~ , 
\end{equation}
where $D_{g}$ and $D_{\gamma}$ are the data sets from the GW and gamma-ray channels, and
($H_{i}$) are the different hypotheses listed above. 

Finally, if we assume complete ignorance about prior probabilities, this expression can be simplified as
\begin{equation}
\Lambda =\frac{I_{\Delta \mathrm{t}}I_{\Omega}}{1 + Q_{g} + Q_{\gamma} + Q_{g}Q_{\gamma}},
\label{eqn:statistics}
\end{equation}
where 
\begin{equation}
    Q_{g} = Q_{g}(D_{g}) = \frac{P(D_{g}|\mathrm{noise})}{P(D_{g}|\mathrm{signal})} 
    \label{eqn:Qg}
\end{equation}
and 
\begin{equation}
    \Qgamma = \Qgamma(D_{\gamma}) = \frac{P(D_{\gamma}|\mathrm{noise})}{P(D_{\gamma}|\mathrm{signal})}
    \label{eqn:Qgamma}
\end{equation} 
are the single-instrument Bayes factors comparing the noise-only and noise+signal hypotheses in the GW and GBM data respectively.
We discuss the calculation of each term in this ranking statistic in Sections~\ref{GBMBF}--\ref{Skyoverlap}. 

Equation~\ref{eqn:statistics} has intuitive limits:
\begin{itemize}
    \item If both signals are very marginal, deep into their noise distribution, then $Q_{g} Q_{\gamma} \gg Q_{g},Q_{\gamma} \gg 1$, and $\Lambda \approx \frac{P(D_g|H_{S})}{P(D_g|H_{N)}} \times \frac{P(D_\gamma|H_{S})}{P(D_\gamma|H_{N)}} \times I_{\Delta \mathrm{t}}I_{\Omega} \ll I_{\Delta \mathrm{t}}I_{\Omega}$. The association is suppressed by both single-instrument Bayes factors, and the parameter overlap must be exceptional for the association to be significant. 
    \item If one signal is very significant (say the GW) then $ Q_{g} \ll 1$ and $\Lambda$ only depends on the significance in the other instrument (and the parameter overlap).
    \item If we have very significant signals in both GW and GBM data, then $ Q_{g} Q_{\gamma} \ll Q_{g}, Q_{\gamma} \ll 1 $, and $ \Lambda \approx I_{\Delta t}I_{\Omega} $: the single-instrument significances become irrelevant, and the only thing that can prove the association between the triggers is the overlap in their inferred parameters.
\end{itemize}

\subsection{GBM Bayes Factor $\Qgamma$}
\label{GBMBF}
Calculating $\Qgamma$ accurately would in principle involve a complete parametric model for the photon flux of an sGRB, for the detailed response of GBM to that photon flux, and for the GBM background.
This would then allow us to write a likelihood function and marginalize it over the relevant parameter space to compute the Bayes factor.
Although attempts have been made to solve parts of this modeling task (see e.g.~\citep{GRMHDNiceMovies}), this is clearly a major challenge.
Instead, here we use an approximate phenomenological model based on a sample of sGRBs observed by GBM and confirmed by other gamma-ray observatories, and a sample of triggers unlikely to be associated with sGRBs.

\begin{figure}
    \centering
     \begin{subfigure}
         \centering
         \includegraphics[width=0.45\textwidth]{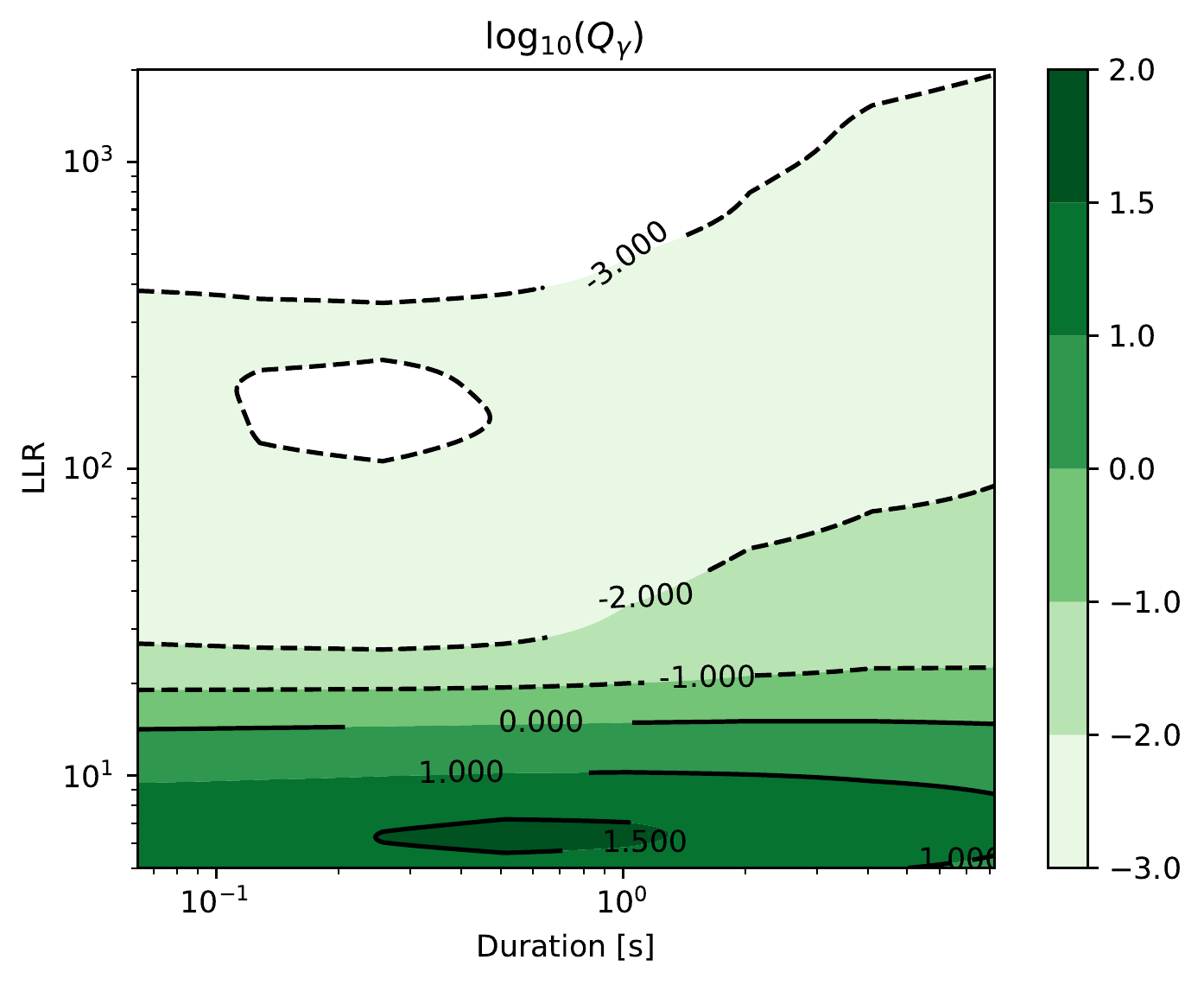}
         \label{fig:2D-GBMkde}
     \end{subfigure}
     \hfill
     \begin{subfigure}
        \centering
        \includegraphics[width=0.45\textwidth]{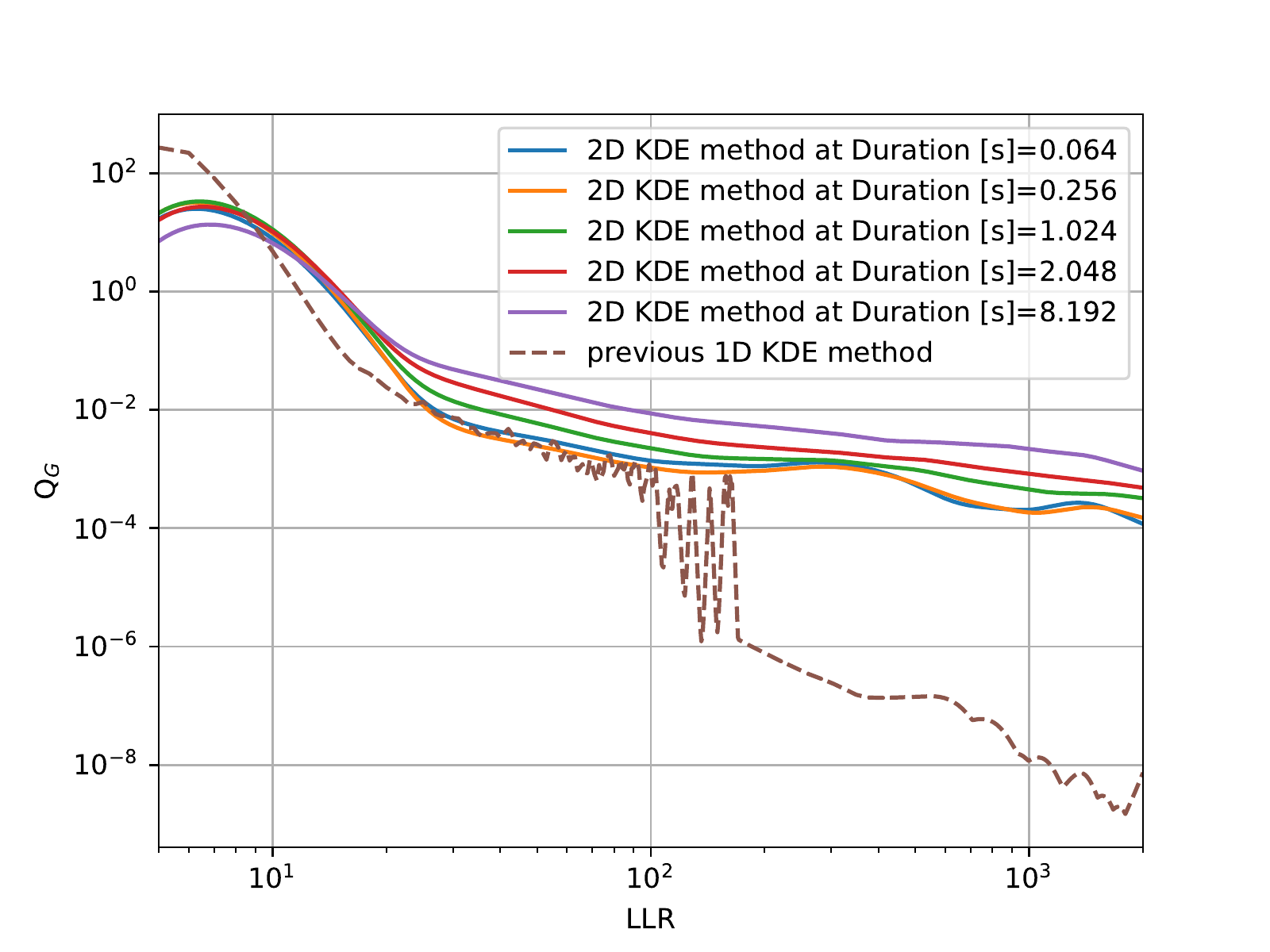}
        \label{fig:comparison_GBM_BF}
     \end{subfigure}
    \caption{GBM Bayes factor computed with a 2D KDE training with real data over $\mathrm{log}_{10}(\mathrm{LLR})$ and $\mathrm{log}_{10}(\mathrm{Duration [s]})$. Top: behavior of the GBM Bayes factor $Q_{\gamma}$ computed with a 2D KDE fitted with real data. Bottom: Comparison between the GBM Bayes factor $Q_{\gamma}$ computed with the 2D KDE (solid lines) and the 1D KDE trained on LLR and used in \citep{Ranking} (dashed line).}
    \label{fig:BFfg}
\end{figure}

Specifically, we approximate $\Qgamma$ following a 2D kernel density estimation (KDE, \citep{KDE}) in the log(duration)--log(LLR) space.
The KDE is fit on a sample of positive triggers. These include a sample of 61 GBM-triggered Targeted Search triggers within 3s of confirmed events (i.e. identified with the GBM untargeted search that has confirmation in other instruments such as INTEGRAL SPI-ACS \citep{Rau_2005} or Swift \citep{2004ApJ...611.1005G}) with $T_{90}<2$ s and $>90\%$ probability of being an sGRB during O2, during the third GW observing run (O3), and between observing runs, and on a negative sample. Conservatively, every non-positive trigger (including detector noise) is considered to be negative.
The negative sample counts 1,897,956 triggers.
The KDE is then evaluated on the GBM triggers and the ratio of the probability functions presented in Equation~\ref{eqn:Qgamma} is computed.
The resulting $Q_{\gamma}$ distribution is shown in the top panel of Figure~\ref{fig:BFfg}.
We can observe that the signal-like region is associated with the highest LLR values.
Moreover, at fixed LLR, triggers with a shorter duration have a smaller (i.e. signal-like) $Q_{\gamma}$ than those with a longer duration.

The bottom panel of Figure~\ref{fig:BFfg} presents the present approximation with an earlier study \citep{Ranking}, which used a simpler 1D KDE on the LLR only, coupled with some assumptions to extrapolate the distributions beyond the range of the available training set.
The 2D KDE method described here produces a similar behavior, but the rapid oscillations in the curve of Bayes factor versus LLR, which were artifacts of the previous method, have disappeared.
We can also see that most of the curves with the 2D KDE Bayes factor are above the 1D KDE curve, meaning that with our 2D KDE-based Bayes factor it is harder for a candidate to be signal-like.

\subsection{GW Bayes Factor $Q_{g}$}
\label{GWBF}

In addition, \texttt{Bayestar} also provides two Bayes factors:
\begin{itemize}
    \item $\mathrm{BSN}$: Bayes factor signal versus noise, compares the probability for a trigger to be signal-like and noise-like. This Bayes factor is qualitatively similar to a S/N and defined as 
    \begin{equation}
        \frac{P(\mathrm{D_g | Signal + Gaussian \ noise})}{P(\mathrm{D_g | Gaussian \ noise})}
    \end{equation}
    \item $\mathrm{BCI}$: Bayes factor coherent versus incoherent, compares the probability of being a coherent signal versus an incoherent signal. It is defined as \begin{equation}
        \frac{P(\mathrm{D_g | C + Gaussian \ noise})}{P(\mathrm{D_g |I + Gaussian \ noise})}
    \end{equation} where $C$ and $I$ are respectively the coherent and incoherent signals. 
\end{itemize}
Intuitively, we expect GW astrophysical signals to have both high $\mathrm{log}_{10}(\mathrm{BSN})$ and $\mathrm{log}_{10}(\mathrm{BCI})$, Gaussian noise to have both small $\mathrm{log}_{10}(\mathrm{BSN})$ and $\mathrm{log}_{10}(\mathrm{BCI})$, and glitches to have a high $\mathrm{log}_{10}(\mathrm{BSN})$ and a small $\mathrm{log}_{10}(\mathrm{BCI})$. \

Two configurations are tested to compute the GW Bayes factor $Q_{g}$. The first one is used in the analysis and the second one is presented in Appendix \ref{appendix:QG}. 
\begin{itemize}
    \item[1.] We use the logarithm of the BCI, which is the result of a model comparison between a coherent GW signal in the entire network and versus a signal that is not coherent (the most likely occurrence being a single-detector signal). 
    \item[2.] We compute a KDE-based GW Bayes factor in the $\mathrm{log}_{10}(\mathrm{BCI})-\mathrm{log}_{10}(\mathrm{BSN})$ plane. 
    We train the KDE on a positive sample composed of GW injections and a background sample made of GW triggers with a FAR above two per day; this is detailed in Appendix~\ref{appendix:QG}.
\end{itemize}

The $\mathrm{BCI}$ is robust to glitches, contrary to the $\mathrm{BSN}$, which is why we only extract the $\mathrm{BCI}$ from the skymaps in the main method presented here. We explore a KDE-based method using both the $\mathrm{BCI}$ and the $\mathrm{BSN}$ in Appendix~\ref{appendix:QG}. Indeed, a KDE requires a training sample, so GW injection skymaps should be generated in addition to skymaps for noise and glitches, which is much more complicated than considering only the $\mathrm{BCI}$.

\subsection{Time Overlap}
\label{Timeoverlap}
The goal of the time offset term $I_{\Delta t}$ is to quantify how probable it is for a pair formed by a GW trigger and a GBM trigger to be separated by a certain amount of time $\Delta t$ = $t_{\mathrm{GBM}}$ -- $t_{\mathrm{GW}}$, where $t_{\mathrm{GW}}$ is the merger time of the GW candidate estimated by the GW searches, and $\mathrm{t}_{\mathrm{GBM}}$ is the central time of the GBM trigger with the maximum LLR.

Here we define the time term as 
\begin{equation}
    I_{\Delta \mathrm{t}} = \left\{
    \begin{array}{ll}
        1 - \frac{|\Delta \mathrm{t}|}{30} & \mbox{if } |\Delta \mathrm{t}| < 30 s, \\
        0 & \mbox{otherwise.}
    \end{array}
\right.
\end{equation}
This is similar to what has been used in \citep{Ranking} under the same assumptions. 
Since we currently have only one joint detection, this choice of prior is essentially arbitrary.
The time delay between GW170817 and GRB 170817A was about $1.7$ s, so we make the prior assumption that the closer in time the two messengers are, the more likely they are to come from the same source.
Our prior extends to much larger delays than $1.7$ s, which, although unlikely according to some prompt emission models, is large enough to allow us to detect so-far-unconfirmed phenomena, such as sGRB precursors (\cite{ShatteringFlares, Stachie:2021uky}).

\subsection{Sky Overlap}
\label{Skyoverlap}
The general form of the Bayes factor for the sky proximity is given in \citep{SKYOVERLAP} as
\begin{equation}
    I_{\Omega} = \int_{}^{} \frac{P(\Omega | \mathrm{D}_g)P(\Omega | \mathrm{D}_\gamma)}{P(\Omega)} d\Omega
    \label{Iomega}
\end{equation}
This sky overlap Bayes factor $I_{\Omega}$ should be large (strongly positive) when the GW and the GBM triggers are overlapping and well localized (small uncertainty region). It will be close to 0 when the two triggers are localized far from each other. The sky overlap is $I_{\Omega} \approx 1$ even if the two triggers are overlapping but the skymaps are uninformative (large uncertainty regions on both sides). In that case the posterior $P(\Omega|D_i)$ (with $i = g \ \mathrm{or} \ \gamma$) is almost equal to the prior $P(\Omega)$ in Equation~\ref{Iomega}, leading to $I_\Omega \approx 1$.  

Moreover, when looking for a CBC producing coincident signals in GW data and Fermi/GBM, we know a priori that it cannot be located behind the Earth. Therefore, a more realistic prior (compared to a uniform prior on the whole sky) for the sky location is zero over the
Earth, and uniform over the portion of the sky not occulted by the Earth. This leads to
\begin{equation}
    P(\Omega)= \left\{
        \begin{array}{ll}
        1/\int_{\bar{\Earth}}^{} d\Omega = 1/f_{vis}4\pi & \mbox{if } \Omega \notin \Earth \\
        0 & \mbox{if } \Omega \in \Earth
        \end{array}
\right.
\end{equation}
where 
\begin{equation}
f_{vis} = \frac{1}{4\pi}\int_{\bar{\Earth}}^{} d\Omega
\end{equation}
is the fraction of the sky not occulted by the Earth. Finally, one can rewrite Equation~\ref{Iomega} as
\begin{equation}
    I_{\Omega}^{EA} = 4\pi f_{vis}\int_{\bar{\Earth}}^{} P(\Omega | \mathrm{D}_g)P(\Omega | \mathrm{D}_\gamma) d\Omega
    \label{IOmegaEA}
\end{equation}
In contrast to an all-sky prior, the Earth-avoiding prior systematically reduces the Bayes factor when the events being compared are well separated from the Earth, 
regardless of how much their skymaps overlap.

\begin{figure}
    \centering
    \includegraphics[width=0.45\textwidth]{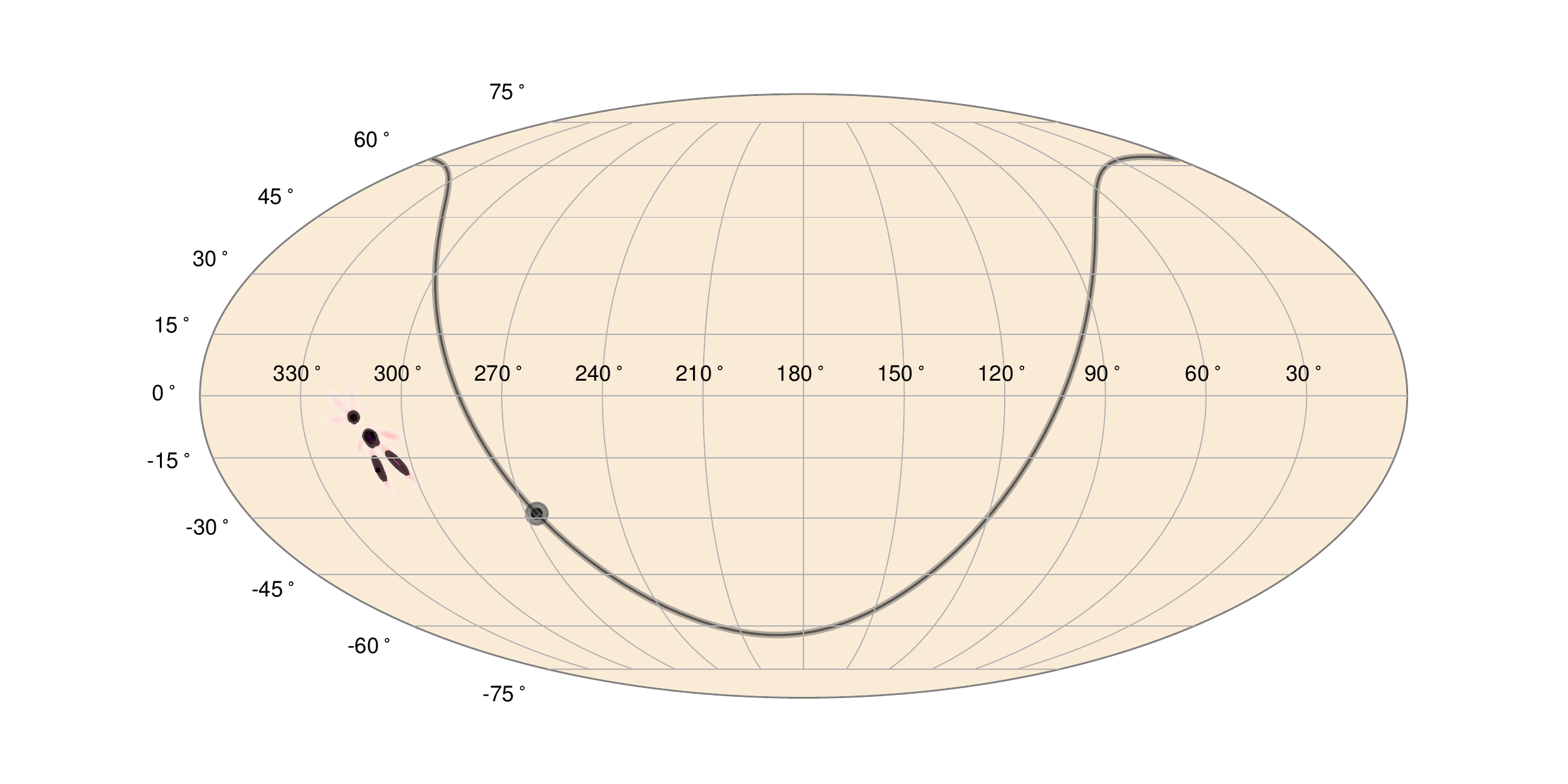}
    \includegraphics[width=0.45\textwidth]{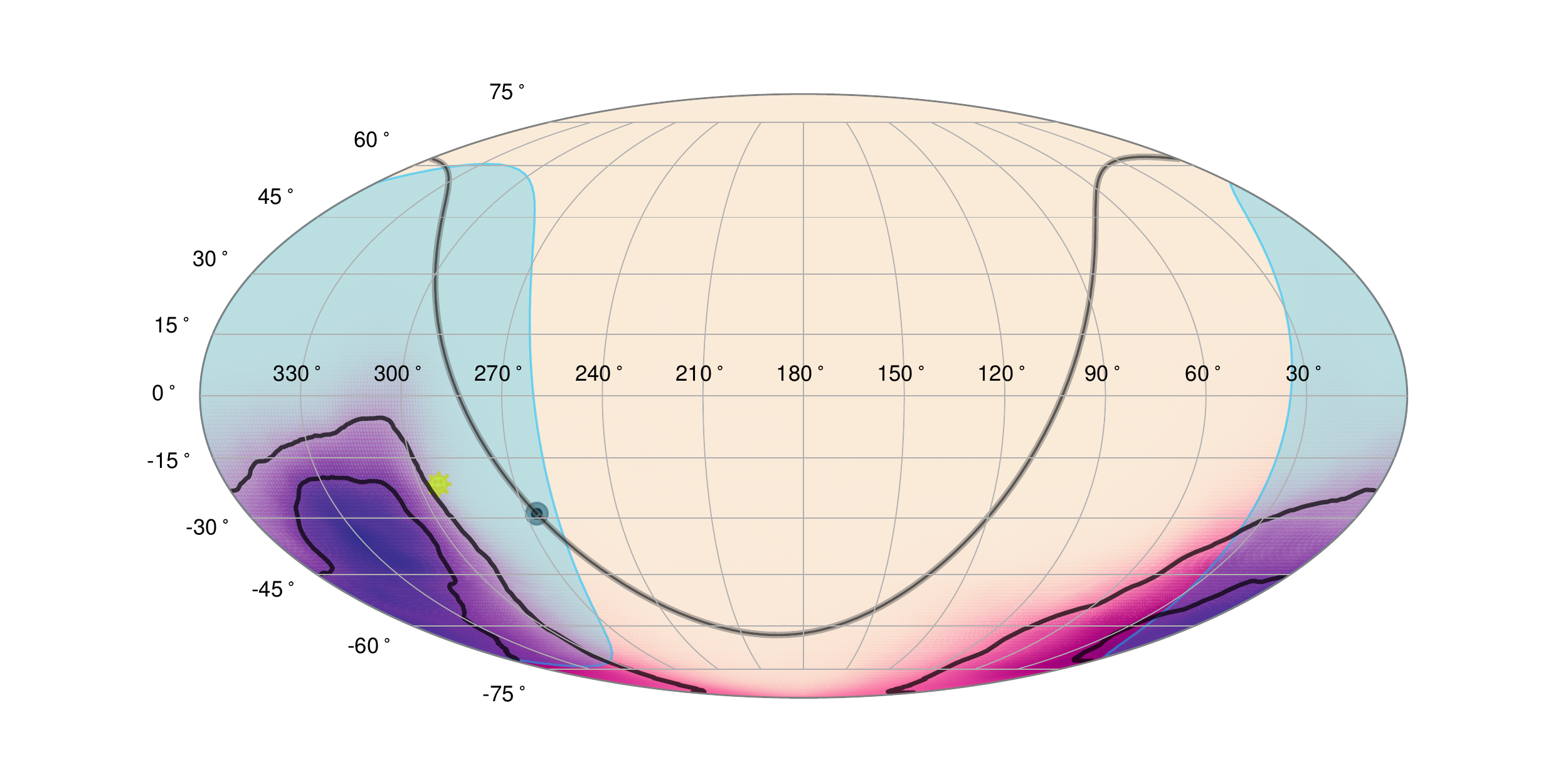}
    \includegraphics[width=0.45\textwidth]{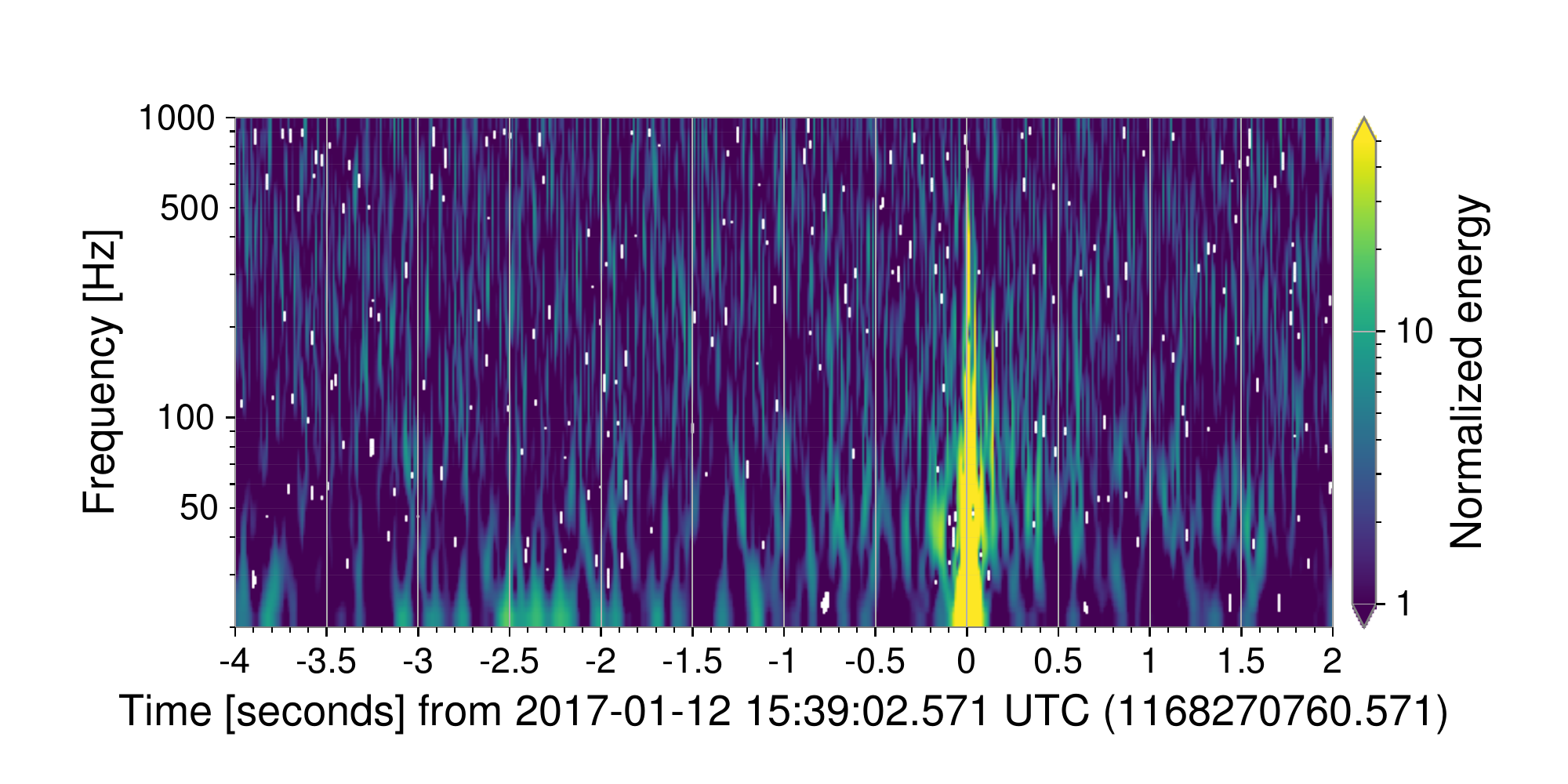}
    \caption{Example of a GW--GBM association that is strongly suppressed by the Earth-avoiding prior. Top: sky localization of the GW trigger. Middle: sky localization of the GBM trigger. The gray solid line represents the Galactic plane, and the Sun is represented by the yellow star in the GBM skymap. The blue region is the Earth's location. The two contour levels represent the 90\% and 50\% credible regions. The darker the purple, the higher the probability. Bottom: spectrogram of L1 data around the GW trigger.}
    \label{fig:skymap}
\end{figure}

When the sky localization of the joint association is most likely behind the Earth, the difference between the sky term with a prior avoiding the Earth and considering a uniform prior over the whole sky (ignoring the presence of the Earth) becomes important. An example is shown in Figure~\ref{fig:skymap}. Here, the morphology seen in the spectrogram in the bottom panel of Figure~\ref{fig:skymap} points out that the GW trigger is unlikely to be a compact binary merger signal but rather a glitch (triggering the high mass templates, $\mathrm{S/N} \approx 73$ and $\chi^2 \approx 259$) and the GBM trigger has a probability of being occulted by the Earth of $83\%$. Hence, it is unlikely to be a GRB but rather noise coming from behind the Earth or from the Earth itself (e.g. Terrestrial Gamma-Ray flashes \citep{2018JGRA..123.4381R}) triggering the detector. This association is highly unlikely to be an astrophysical association and should be suppressed by its sky overlap term. When we set to zero the probability behind the Earth, the Earth-avoiding sky term is $I_{\Omega}^{EA} \approx 3.08\times10^{-10}$, which indicates that the association is more likely to be an accidental coincidence. In contrast, with the all-sky prior, it becomes $I_{\Omega} \approx 4.06$. Since both triggers have most of their posterior distribution behind the Earth, the sky overlap term of the association should be very close to zero, which is the case with $I_{\Omega}^{EA}$. The discrete computation of the Earth-avoiding prior is detailed in Appendix~\ref{appendix:skymap}.

\subsection{Calculation of the Significance}
Eventually, we want to assign a statistical significance to each foreground pair. To do this, we need to compare the foreground to the background. The FAR (or the inverse false alarm rate, IFAR [yr]) of a candidate gives a quantitative idea of how often noise generates a candidate. It is defined as the rate of background with ranking statistic equal to or higher than the one observed for the candidate in question \citep{PyCBC, Morr_s_2023}. Therefore, it gives a frequentist interpretation to the ranking statistic.
We want to generate the foreground sample, i.e, identify pairs of GW--GBM triggers that could come from a common astrophysical source, sort these pairs with our ranking statistic, and compute their FAR.
A background sample is needed to assign it to each foreground event. The same set of triggers is used to generate both the foreground and the background.
To compute the background sample, we use a time-slide method \citep{W_s_2009} in which we time-shift the GBM triggers by a predetermined offset, and we look for (fake) coincidences between the GW triggers and the time-shifted GBM triggers. 
To choose an optimal background interval (i.e. the number of time slides and the time difference between two time slides) one must consider some requirements. First, keep the interval to a minimum in order to have \textquote{local} estimates of the background, so the detector is in the same state during the background time interval to avoid bias. Second, the interval chosen should not be extremely large, to avoid a high computational cost. Third, the interval needs to be large enough to reach an interesting FAR ($<$1/[1000 yr]) for claiming a discovery. 
All these requirements lead us to start with a shorter time interval of $\pm5\times 10^4$ s ($\sim$27 hours around the GW trigger) in 100 s steps for testing, and then use the larger interval from $-1.80070\times 10^5$ s to $+1.8\times 10^5$ s ($\sim$4 days around the GW trigger) in 70 s steps for final estimates of significance.
The lack of symmetry between the lower and upper boundaries is chosen for computational reasons.

The time offset chosen to compute the background must be greater than twice the maximum time offset of 30s, considered in this analysis to be a nonphysical time delay between a CBC and any possible GRB emission resulting from it. We use a $\pm$70 s offset to accumulate background associations.
 We repeat this process multiple times, each with a different nonzero integer multiple of 70s, and accumulate the background distribution of $\Lambda$ values.
For the foreground, we rank the pairs with the same statistic but without time-shifted GBM triggers in order to find the potential GBM--GW candidate pairs.

\section{Configurations and Results}
\label{results}
To check the validity of our method against GW170817--GRB 170817A we apply it to the data from O2. We first present in Section~\ref{Preliminary} the results obtained using the naive time ranking statistic described in Section\ref{RankingStat} and computing the background associations with time slides from $-5\times 10^4$ to $+5\times 10^4$ with steps of 100 s. We then focus on the main analysis of this paper using the Bayesian ranking statistic defined above. The background associations are computed by shifting with time slides from $-1.80070\times 10^5$ to $+1.8\times 10^5$ with steps of 70 s and several configurations are tested to maximize the significance of GW170817--GRB 170817A:

\begin{itemize}
    \item[1.] In Section~\ref{separation} we present configuration $n^{\circ}1$ consisting in the computation of the significance of the foreground associations by separating the associations by GBM spectral value and GBM duration. The Targeted Search uses with three spectral models \citep{Goldstein_2020}: a \textquote{soft} Band function \citep{1993ApJ...413..281B}, a \textquote{normal} Band function , and a \textquote{hard} exponentially cutoff power law. By separating by spectral value and duration we will compare GW170817--GRB 170817A to associations with the same characteristics on the GBM trigger side. This prevents us from comparing GW170817--GRB 170817A to loud associations whose properties correspond less to GRBs coming from neutron star mergers. So we treat each different (spectral value--duration) pair as an independent search and we apply a final trial factor to the FARs to account for the number of searches.
    \item[2.] Secondly, we describe in Section~\ref{remove-spec-dur} the computation of the significance of the foreground associations without the separation by GBM spectral value and GBM duration (configuration  $n^{\circ}2$).
    \item[3.] Finally, in Section~\ref{preselection} we discuss the configuration $n^{\circ}3$: no separation by GBM spectral value and GBM duration and application of a preliminary cut of the GW triggers based on the FAR. We remove the triggers with a FAR above 2 per day, threshold inspired by GWTC-3 \citep{LIGOScientific:2021djp}.
\end{itemize}

\subsection{Naive Ranking Statistics}
\label{Preliminary}
We first run this analysis on O2 data with the naive time and naive time and sky ranking statistics summarized by Equation~\ref{eqn:naivetime} and Equation~\ref{eqn:naivetimesky}. Background associations are computed using time shifts going from -50,000 to +50,000 s with a step of 100s. The results of the ranking statistic including only the time proximity and the one including the time and sky proximity are very similar, so we only present the naive time ranking statistic. 

\begin{figure*}
    \centering
    \includegraphics[width=0.96\textwidth]{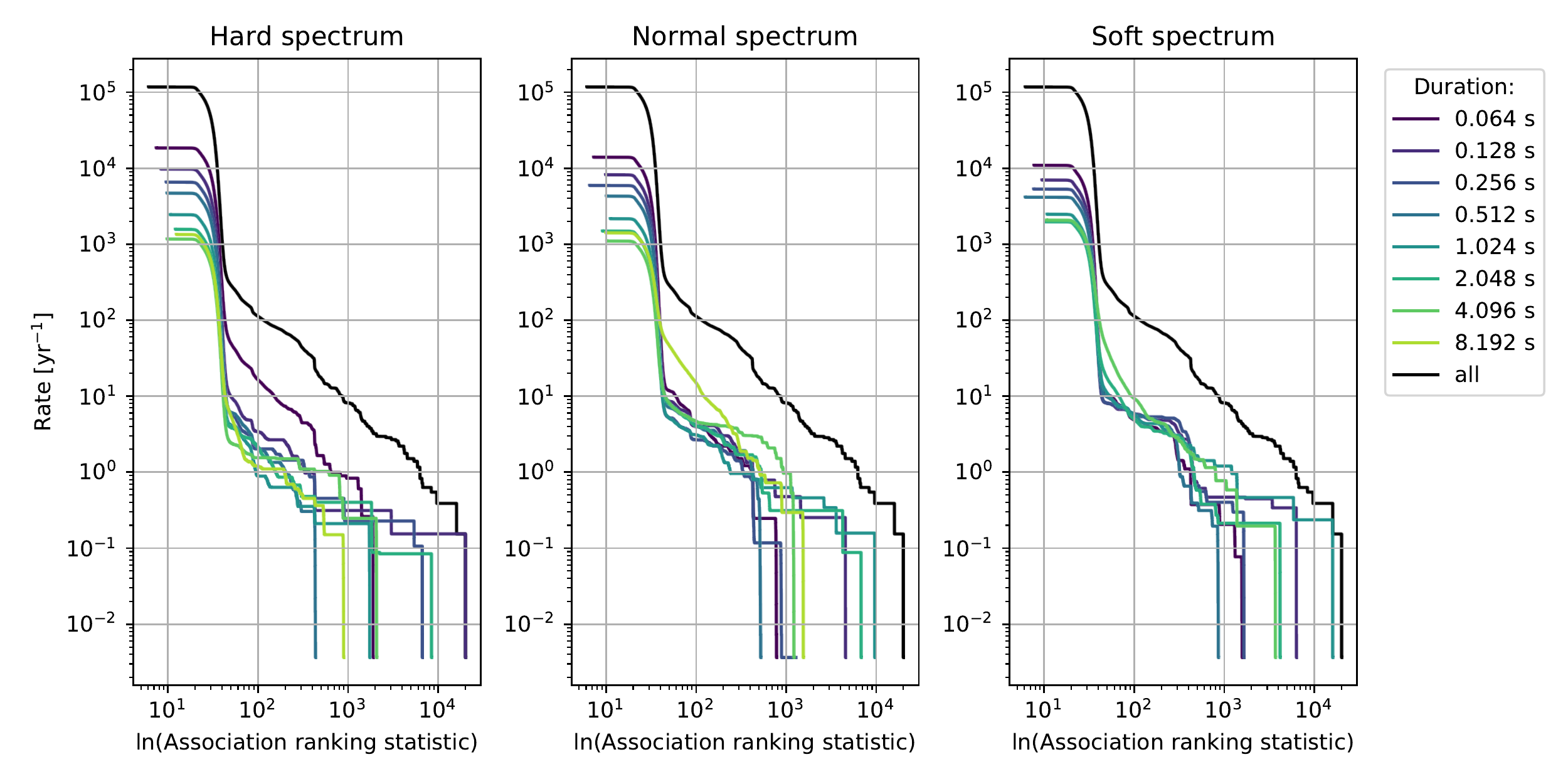}
    \caption{Background rate as a function of the naive time ranking statistic. Left: associations with a hard GBM spectrum. Middle: associations with a normal GBM spectrum. Right: associations with a soft GBM spectrum. The black curve represents all the associations regardless of the duration and spectral hardness of their GBM triggers.}
    \label{fig:naivetime}
\end{figure*}

\begin{figure}
     \centering
     \begin{subfigure}
         \centering
         \includegraphics[width=\columnwidth]{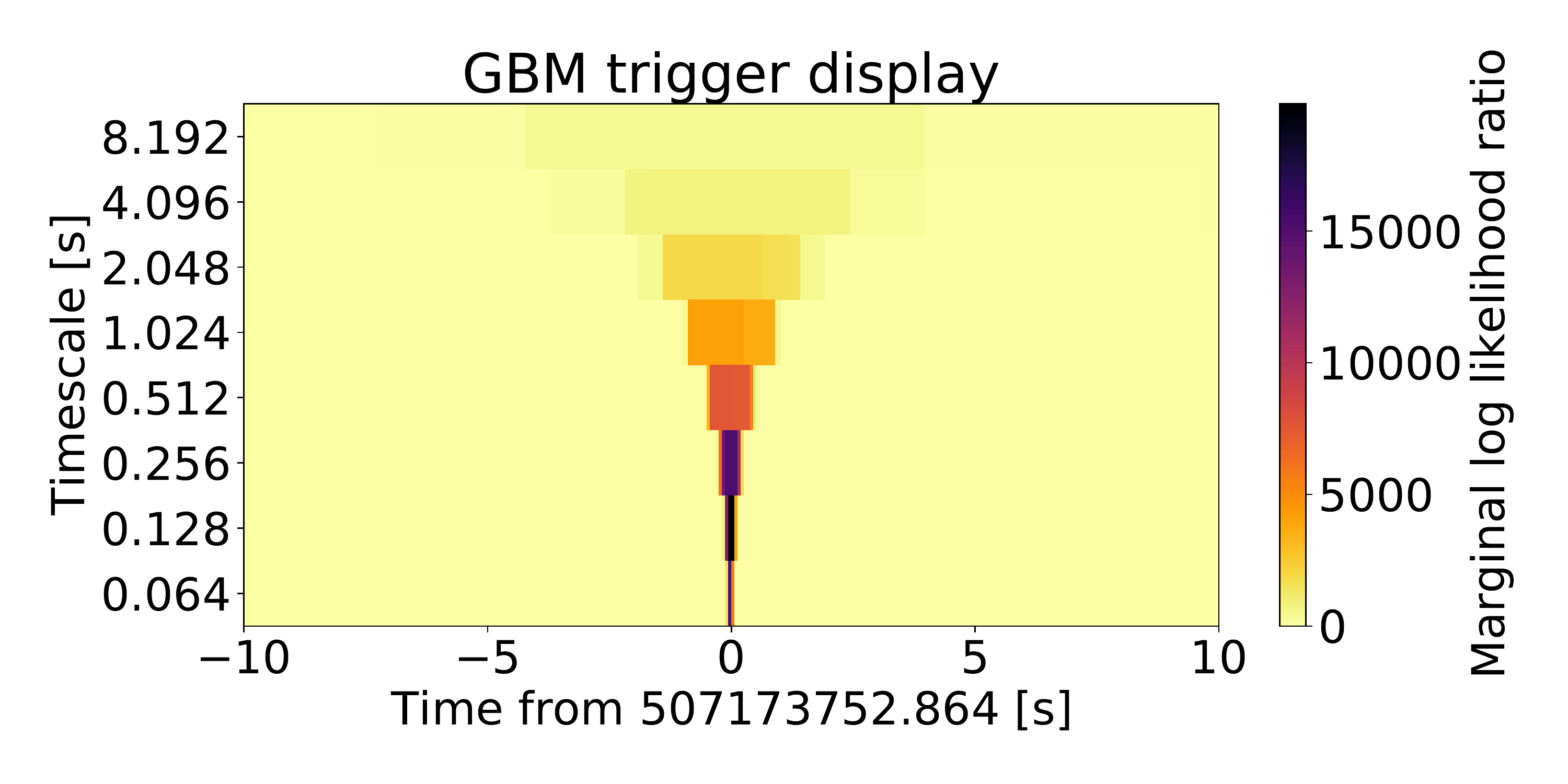}
         \label{fig:waterfallnaive}
     \end{subfigure}
     \hfill
     \begin{subfigure}
         \centering
         \includegraphics[width=\columnwidth]{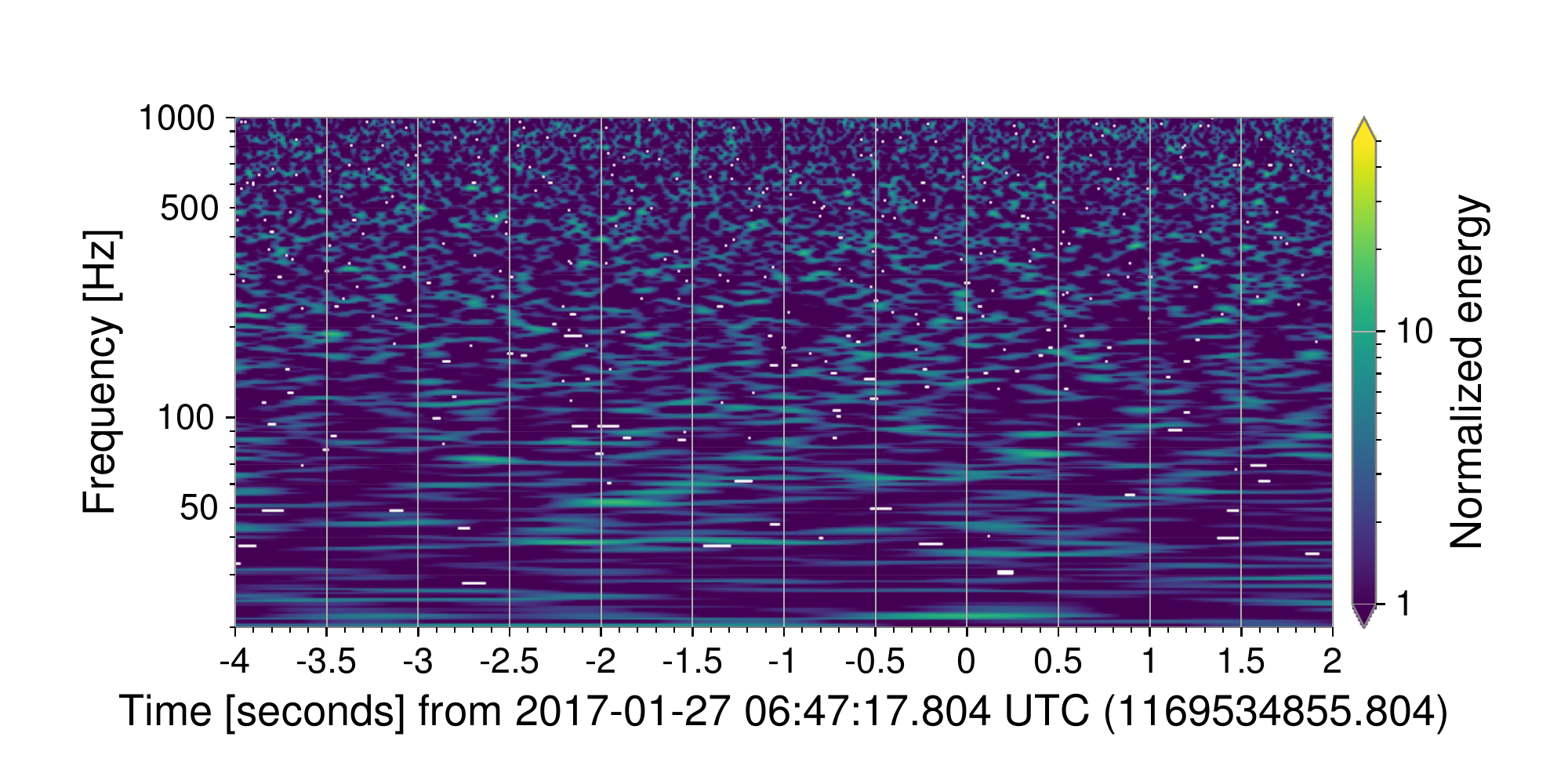}
         \label{fig:H1naive}
     \end{subfigure}
     \hfill
     \begin{subfigure}
         \centering
         \includegraphics[width=\columnwidth]{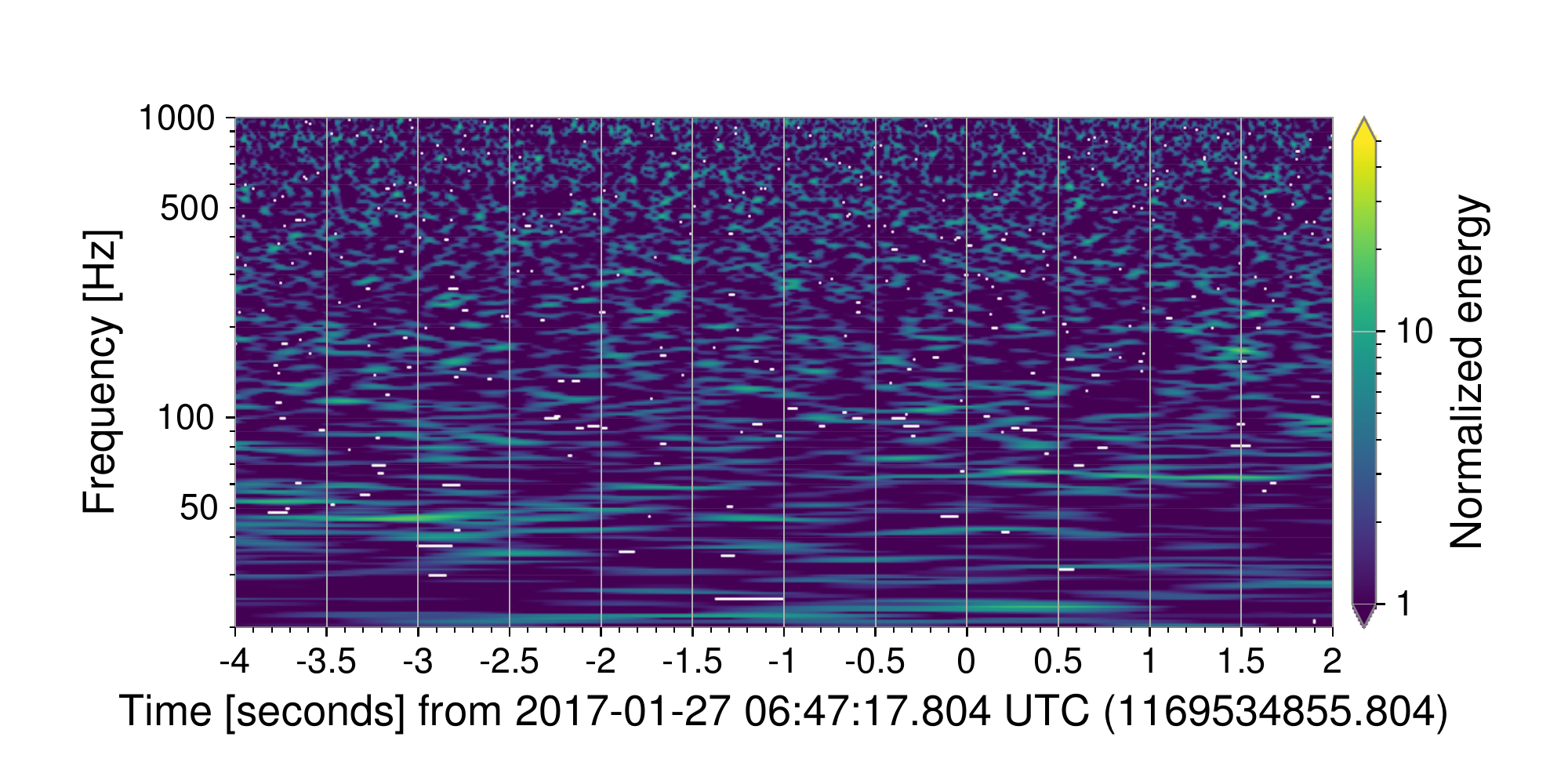}
         \label{fig:L1naive}
     \end{subfigure}
     \caption{Display of the first most significant background association with the naive time statistic. The GBM trigger here is related to GRB 170127C and no GW confident event has been found at the GPS time of the GW trigger.
     Top: Waterfall plot representing the GBM trigger. Middle: spectrogram representing the GW trigger in the H1 detector. Bottom: Spectrogram representing the GW trigger in the L1 detector.}
     \label{fig:naivedisplay}
\end{figure}

Figure~\ref{fig:naivetime} represents how the background behaves. The associations are shown separated by GBM spectral value (the three panels) and GBM duration (the colored curves). Here, the rate of some curves is an order of magnitude smaller than the rate of others. In the right panel of Figure~\ref{fig:naivetime} the curve representing the associations containing an 8.192 s-soft GBM trigger (lightest green curve) does not appear. The GBM Targeted Search intentionally removed such GBM triggers because they contaminate the background. In these plots we can see that the background associations go to extremely high values of association rank.
Indeed, as shown in Table~\ref{table:preliminary-results} and Figure~\ref{fig:naivedisplay}, the naive time and naive time and sky statistics are limited by random coincidences between actual bright short GRBs and random noise in the GW detectors. Moreover, we can see that adding the sky overlap here would not affect the results because of the extremely high values Since the GBM LLR for bright events is several orders of magnitude larger than any of the other quantities in the ranking statistic, the ranking statistic just reduces to the LLR for bright GRBs and the other properties become irrelevant. Therefore, the naive statistics cannot be used unless the various quantities summed together have similar magnitudes.

\begin{table*}
\centering
 \begin{tabular}{|c |c |c |c |c |c |c |c |c |} 
 \hline
 \multicolumn{1}{|c|}{} & \multicolumn{2}{|c|}{GW Properties} & \multicolumn{3}{|c|}{GBM Properties} & \multicolumn{3}{|c|}{Joint Properties} \\
 \hline
Rank & Merger Time & $\hat{\rho}_g$ & Duration (s) & Spectrum & LLR & $\Delta t$ (s) & Time Shift (s) & ln($\Lambda$)\\ [0.5ex] 
 \hline
1 & 1,169,534,855.804 & 7.694 & 0.128 & hard & 19,827.204 & 10.060 & 18,700
& 19,856.394\\ 
2 & 1,169,552,972.042 & 7.602 & 0.128 & hard & 19,827.204 & -6.178 & 36,800 & 19,855.867 \\ 
3 & 1,169,529,460.304 & 7.406 & 0.128 & hard & 19,827.204 & 5.560 & 13,300 & 19,854.422 \\
4 & 1,169,558,255.479 & 7.406 & 0.128 & hard & 19,827.204 & 10.385 & 42,100 & 19,854.202 \\ [1ex] 
 \hline
 \end{tabular}
  \caption{Properties of the first four most significant background associations with the naive time statistic. $\hat{\rho}_g$: Network reweighted SNR. $\Delta t$: time delay between the GBM and the GW trigger. ln($\Lambda$): natural logarithm of the association rank value.}
 \label{table:preliminary-results}
\end{table*}

Finally, if one computes $\Lambda$ in the case of GW170817--GRB 170817 with the naive time statistic, a value of ln$(\Lambda) \approx$ 580 is found, which will not give a significant FAR compared to the most significant background associations from Table~\ref{table:preliminary-results} (with ln$(\Lambda) \approx$ 19,856) that are contaminated by extremely loud triggers in the GBM channel.

\subsection{Bayesian Ranking Statistic and Separation of the Associations by GBM Spectral Value and GBM Duration}
\label{separation}

We now switch to the Bayesian statistic (Equation~\ref{eqn:statistics}) and present the background behavior, then focus on the top background and foreground associations.

\subsubsection{Background Associations}
\begin{figure*}
    \centering
    \includegraphics[width=0.96\textwidth]{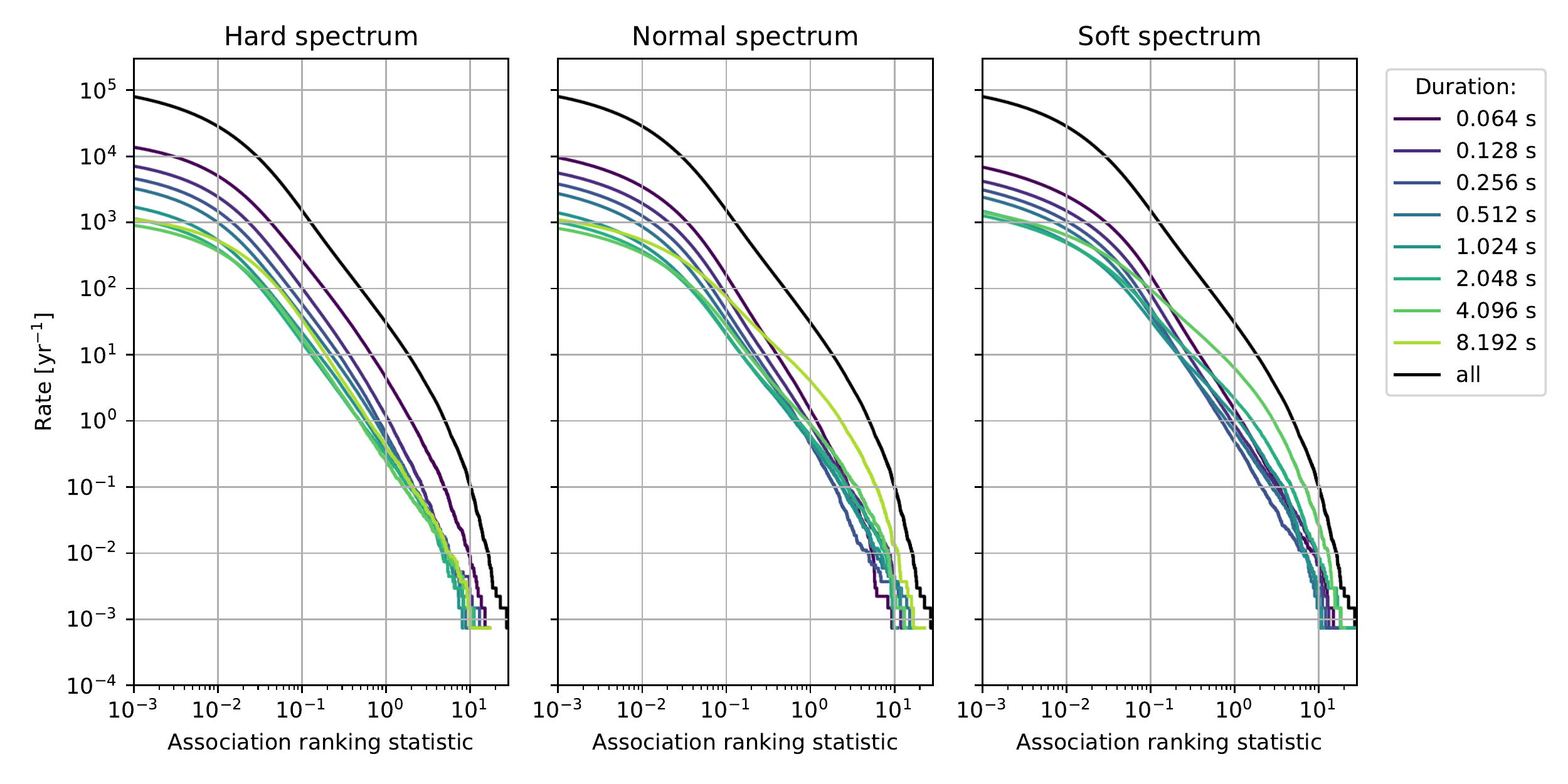}
    \caption{Background rate as a function of the Bayesian ranking statistic (configuration 1). Left: associations with a hard GBM spectrum. Middle: associations with a normal GBM spectrum. Right: associations with a soft GBM spectrum. The black curve represents all the associations regardless of the duration and spectral hardness of their GBM triggers.}
    \label{fig:background_tuning1}
\end{figure*}

\begin{figure*}
    \centering 
    \includegraphics[width=0.45\textwidth]{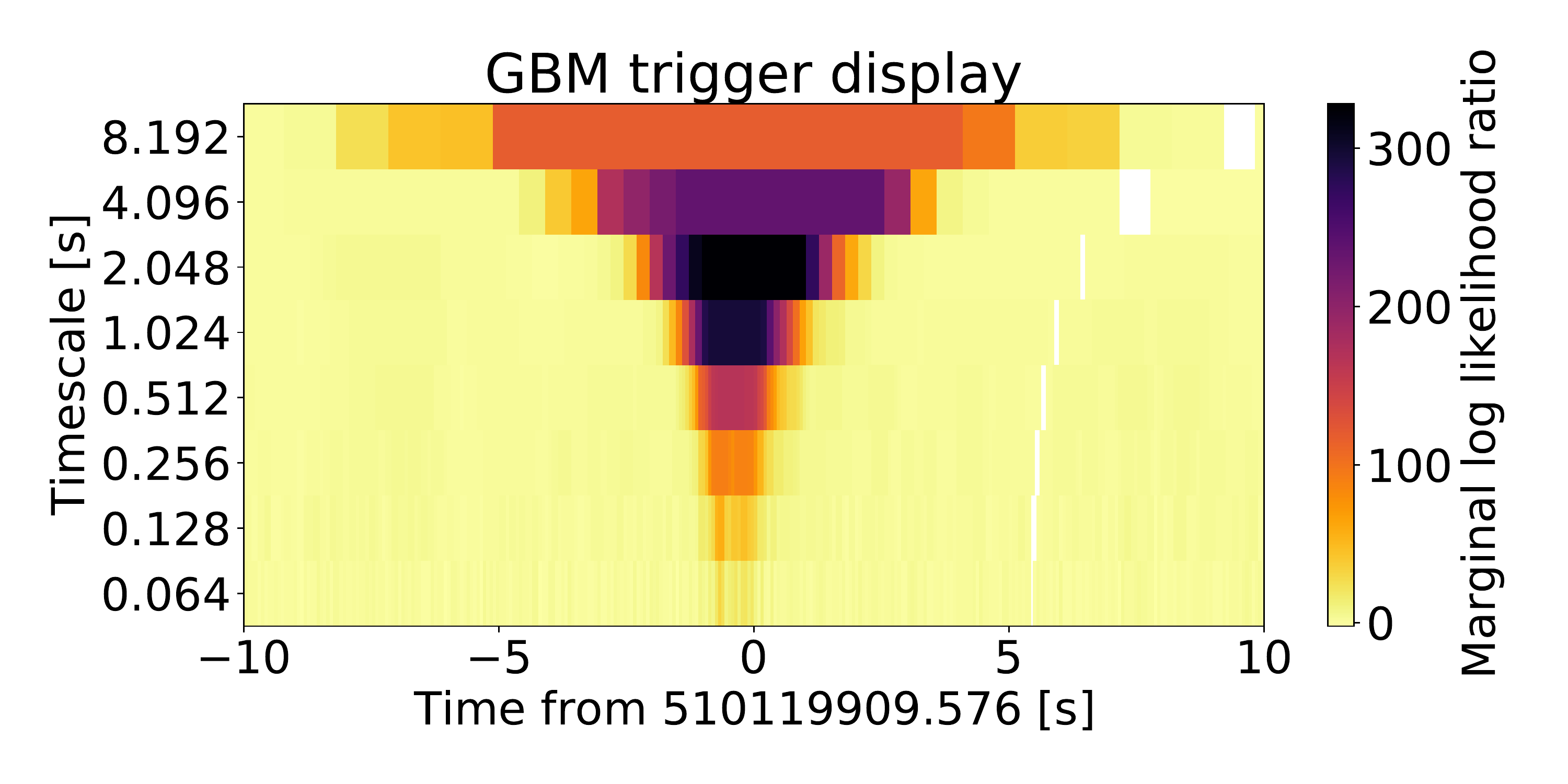}
    \includegraphics[width=0.45\textwidth]{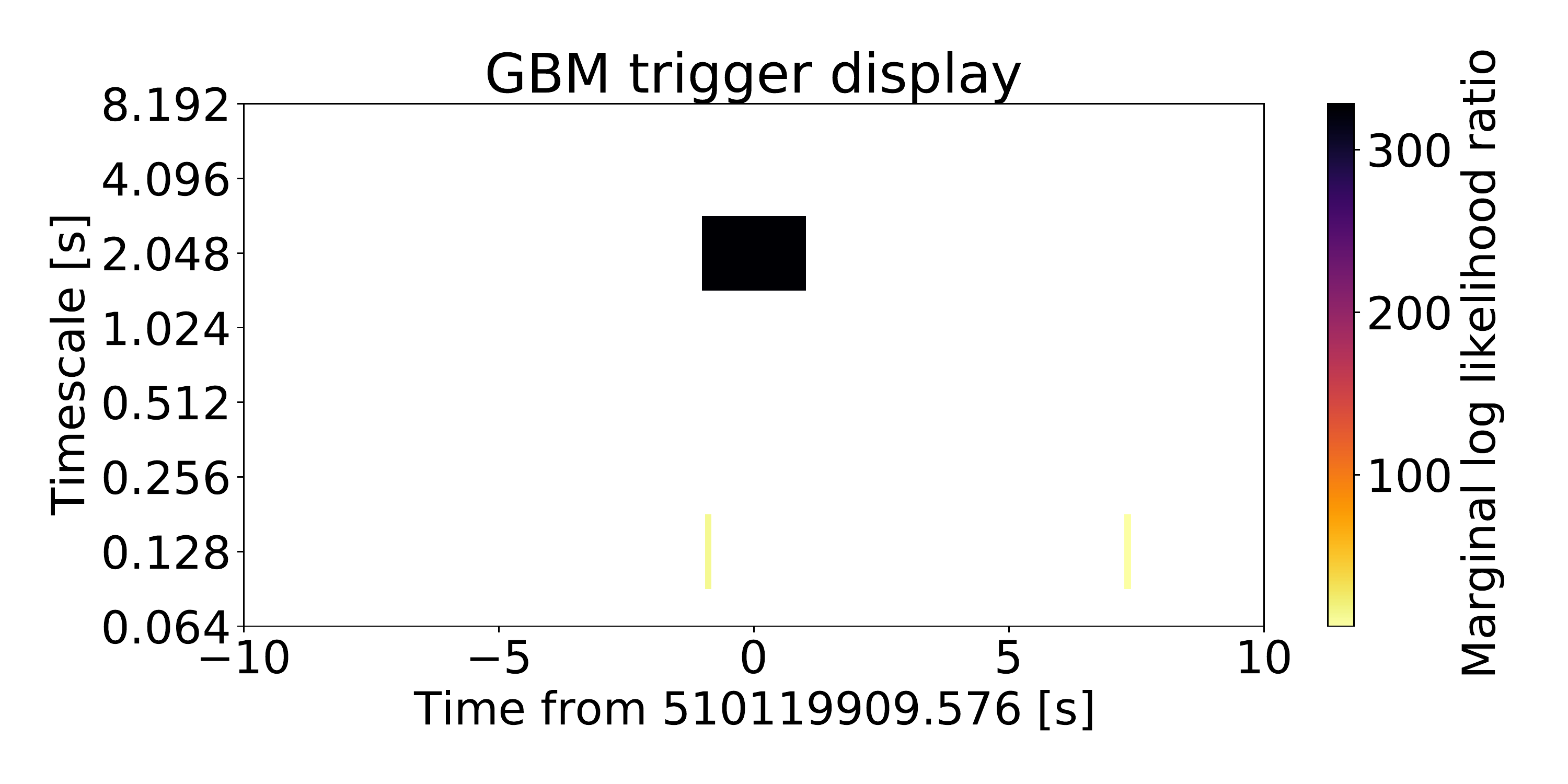}
    \includegraphics[width=0.45\textwidth]{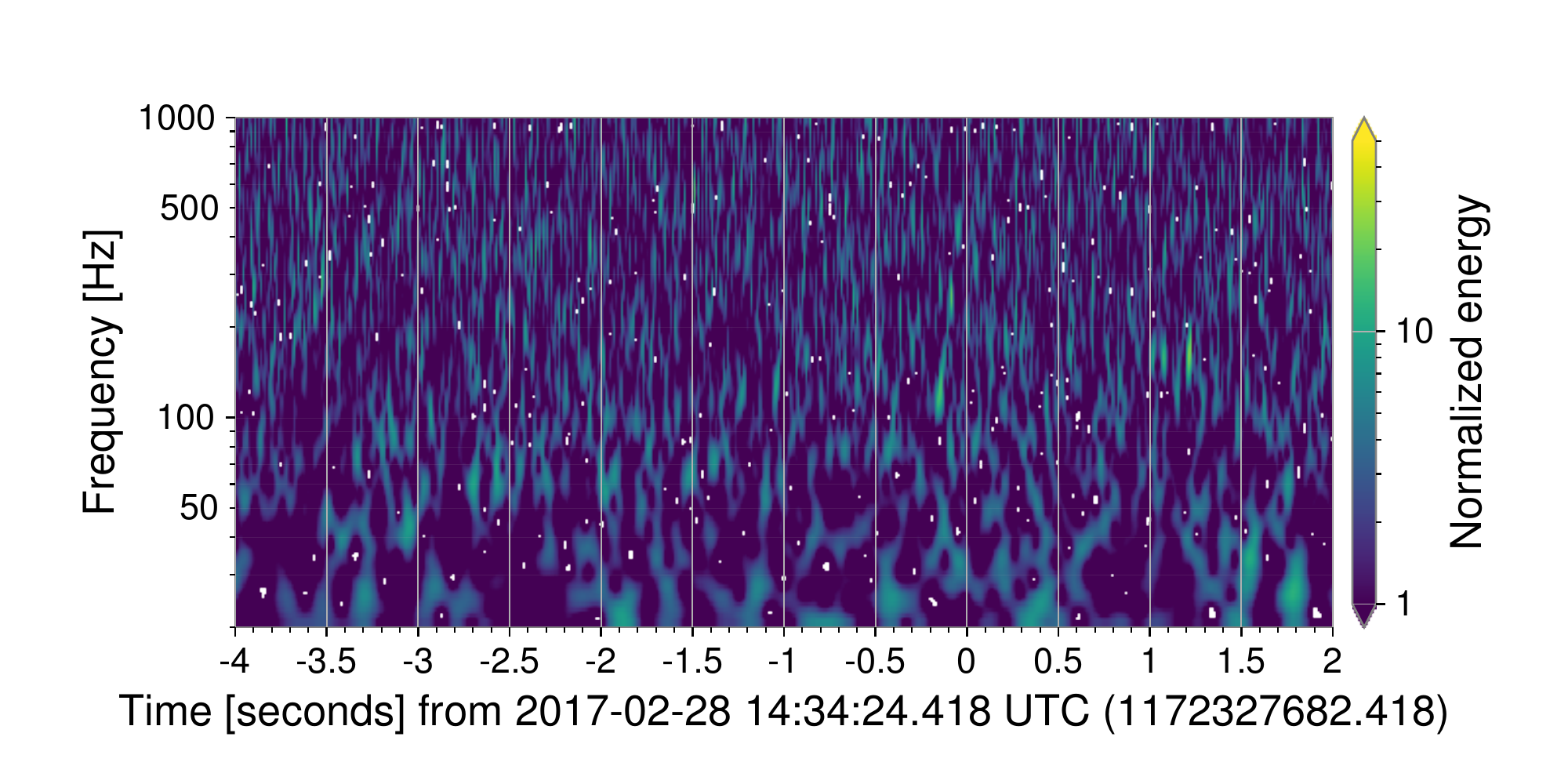}
    \includegraphics[width=0.45\textwidth]{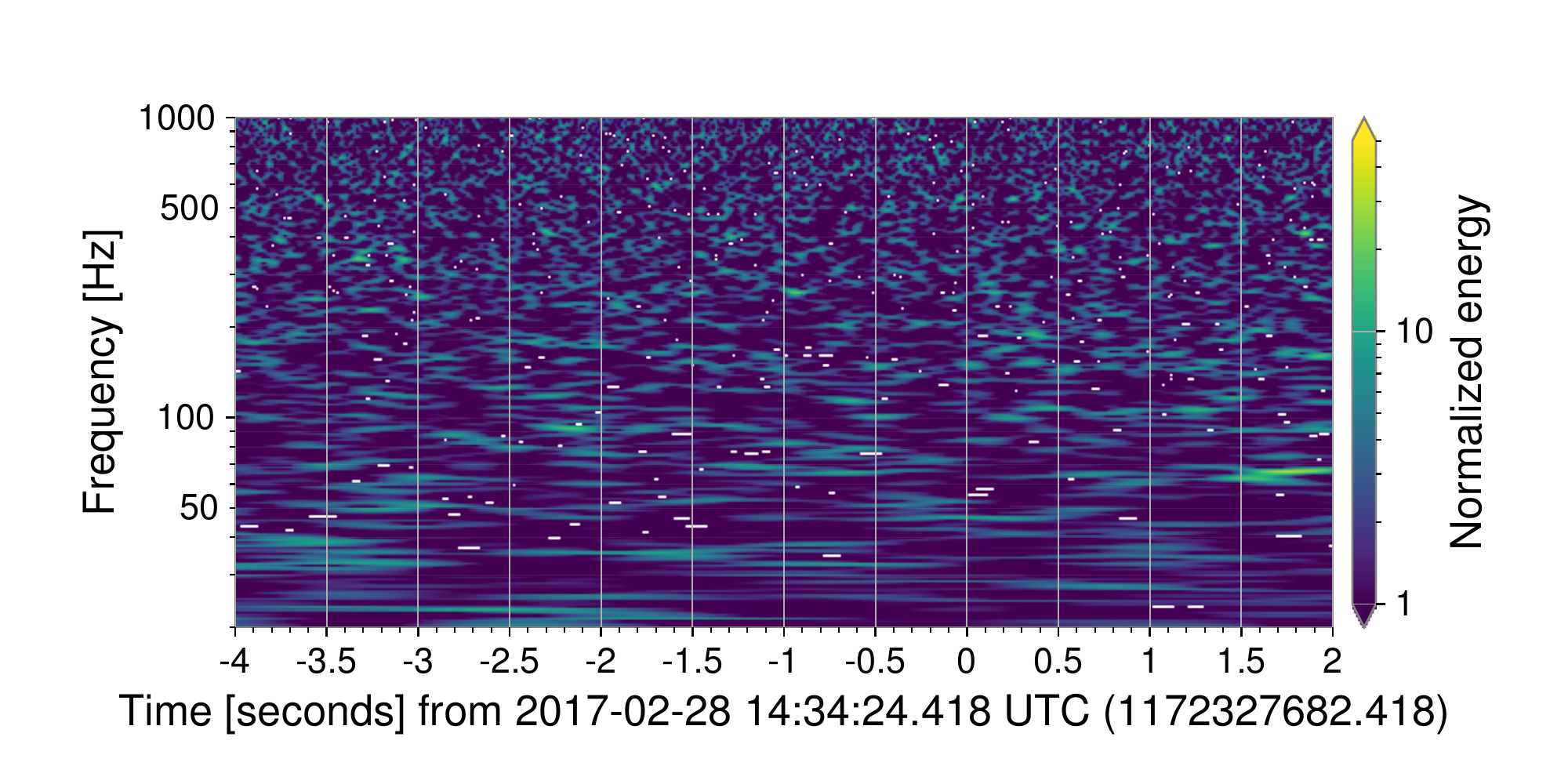}
    \includegraphics[width=0.45\textwidth]{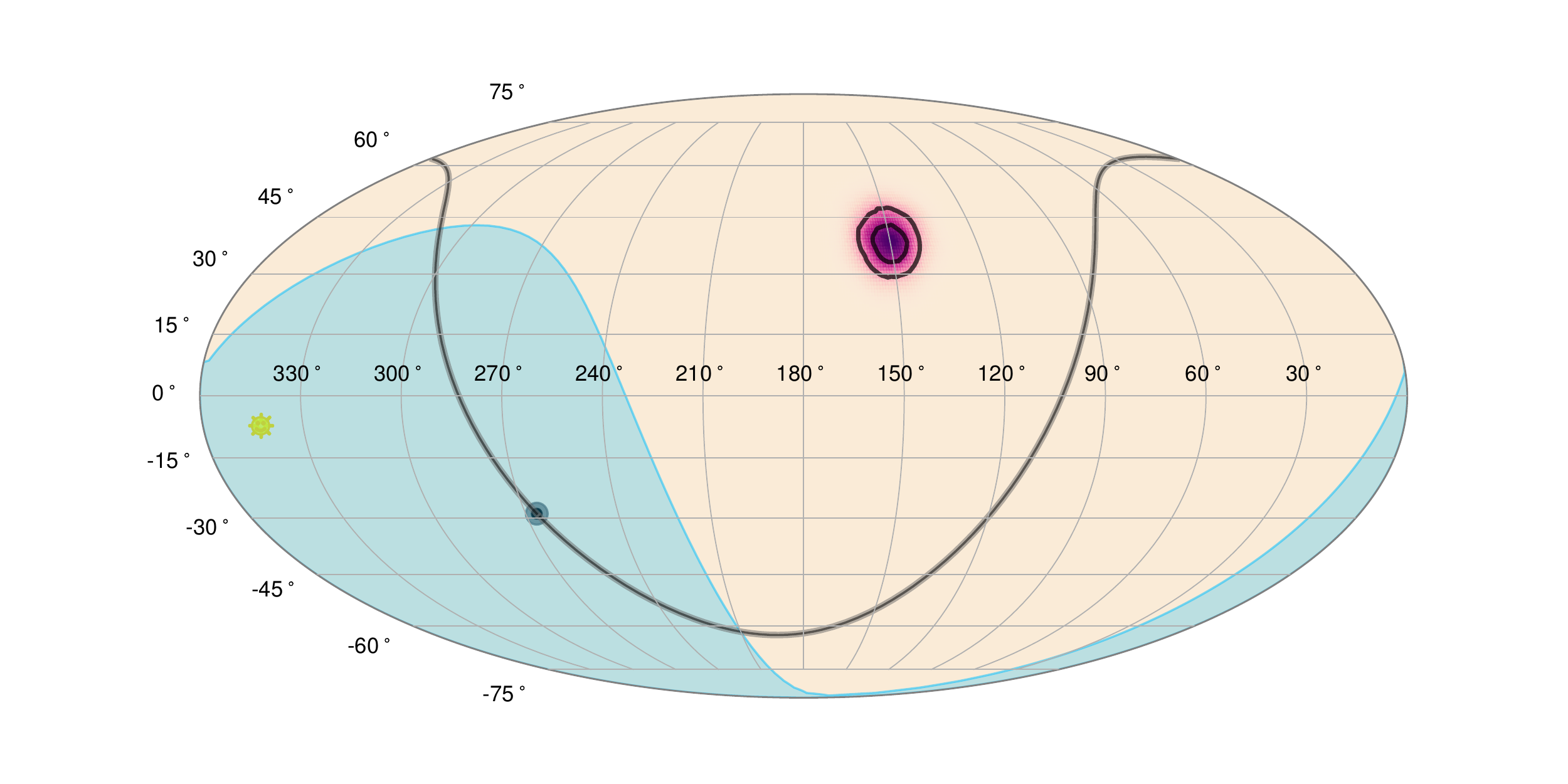}
    \includegraphics[width=0.45\textwidth]{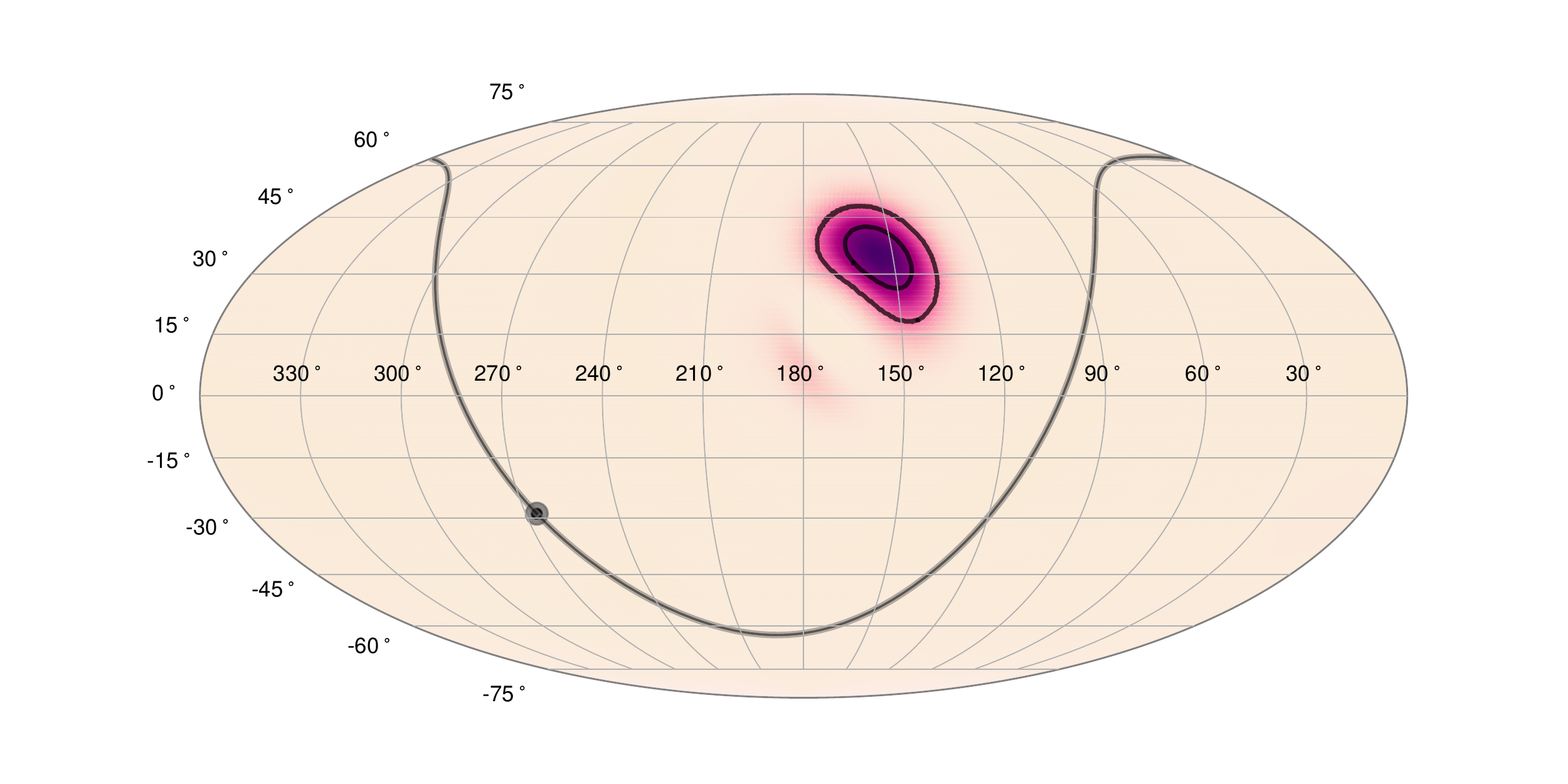}

    \caption{Display of the most significant background association with configuration 1. Top: GBM trigger display with waterfall plots of the GBM trigger (left) and clustered GBM trigger (right). Middle: GW trigger display (Left: H1 spectrogram, Right: L1 spectrogram) Bottom: Sky localization of the association showing the GBM trigger (left) and GW trigger (right). The gray solid line represents the Galactic plane and the yellow star in the GBM skymap represents the Sun. The blue region is the Earth's location. The two contour levels represent the 90\% and 50\% credible regions. The darker the purple, the higher the probability.}
    \label{fig:display-background-config1}
\end{figure*}

Figure~\ref{fig:background_tuning1} represents how the statistic behaves for the background associations. The associations are separated by GBM spectral value (the three panels) and duration. As for the naive time ranking statistic (Figure~\ref{fig:naivetime}), the rate of some curves is an order of magnitude smaller than the rate of others. It goes above 1 only for a small rate with a maximum association rank of $\sim$27 at a background rate of almost $10^{-3} \mathrm{yr^{-1}}$; in other words, we would expect one fake (background) association to have such an association rank per $10^{3}$ yr. Otherwise, most of the background has an association rank smaller than 1. Table~\ref{table:tuning1_background} shows the properties of the first four top ranked background associations and Figure~\ref{fig:display-background-config1} displays the first one. In the first row, in the GBM channel, the candidate has a signal-like $Q_{\gamma}$. It is reported on the Gamma-ray Coordinates Network (GCN; \citep{2000AIPC..526..731B}) as GRB $n^\circ$510119909 \citep{510119909}. On the GW side, the candidate has a signal-like $Q_{g}$ but it is more likely to be noise when looking at the H1 and L1 spectrograms and considering that we did not find any GW confident event at the GPS time of this trigger.
More generally, the top background comprises associations with signal-like GBM candidates and GW triggers that are more likely to be noise. It is composed of very diverse GBM triggers, bright and less bright GBM triggers (large and small LLR), a large range of duration (going from 0.064s to 8.192s), and all spectral values (hard, normal, and soft). This diversity reassures us that they are accidental coincidences.

\begin{table*}
\centering
 \begin{tabular}{|c |c |c |c |c |c |c |c |c |c |c |} 
 \hline
 \multicolumn{1}{|c|}{} & \multicolumn{2}{|c|}{GW Properties} & \multicolumn{4}{|c|}{GBM Properties} & \multicolumn{4}{|c|}{Joint Properties} \\
 \hline
Rank & Merger Time & $Q_g$ & Duration (s) & Spectrum & LLR  & $Q_{\gamma}$ & $\Delta t$ (s) & $I^{EA}_{\Omega}$ & Time shift (s) & $\Lambda$\\ [0.5ex] 
 \hline
1 & 1,172,327,682.418 & $7.47\times10^{-1}$
 & 2.048 & soft & 328.1 & $1.84\times10^{-3}$ & 0.158
& $48.4$ & -134,640 & $27.5$\\ 
2 & 1,177,306,450.914 & $1.13\times10^{-1}$ & 0.256 & soft &  1,595.4 & $2.15\times10^{-4}$ & 13.854 & $55.4$ & 22,650 & $26.8$ \\
3 & 1,165,069,774.280 & $2.05\times10^{-1}$ & 8.192 & normal & 64.28 & $1.53\times10^{-2}$ & 1.216 & $28.8$ & -104,540 & $22.6$ \\ 
4 & 1,165,070,949.873 & $1.63\times10^{-1}$ & 4.096 & soft & 27.66 & $5.23\times10^{-2}$ & 0.679 & $25.2$ & -16,410 & $20.1$ \\ [1ex] 
 \hline
 \end{tabular}
  \caption{Properties of the first four most significant background associations with configuration 1. $Q_g$: GW Bayes Factor. $Q_{\gamma}$: GBM Bayes Factor. $\Delta t$: time delay between the GBM and the GW trigger. $I^{EA}_{\Omega}$: sky overlap value. $\Lambda$: association rank value.}
 \label{table:tuning1_background}
\end{table*}

\subsubsection{Foreground Associations and Significance}
\begin{figure}
    \centering
    \includegraphics[width=0.45\textwidth]{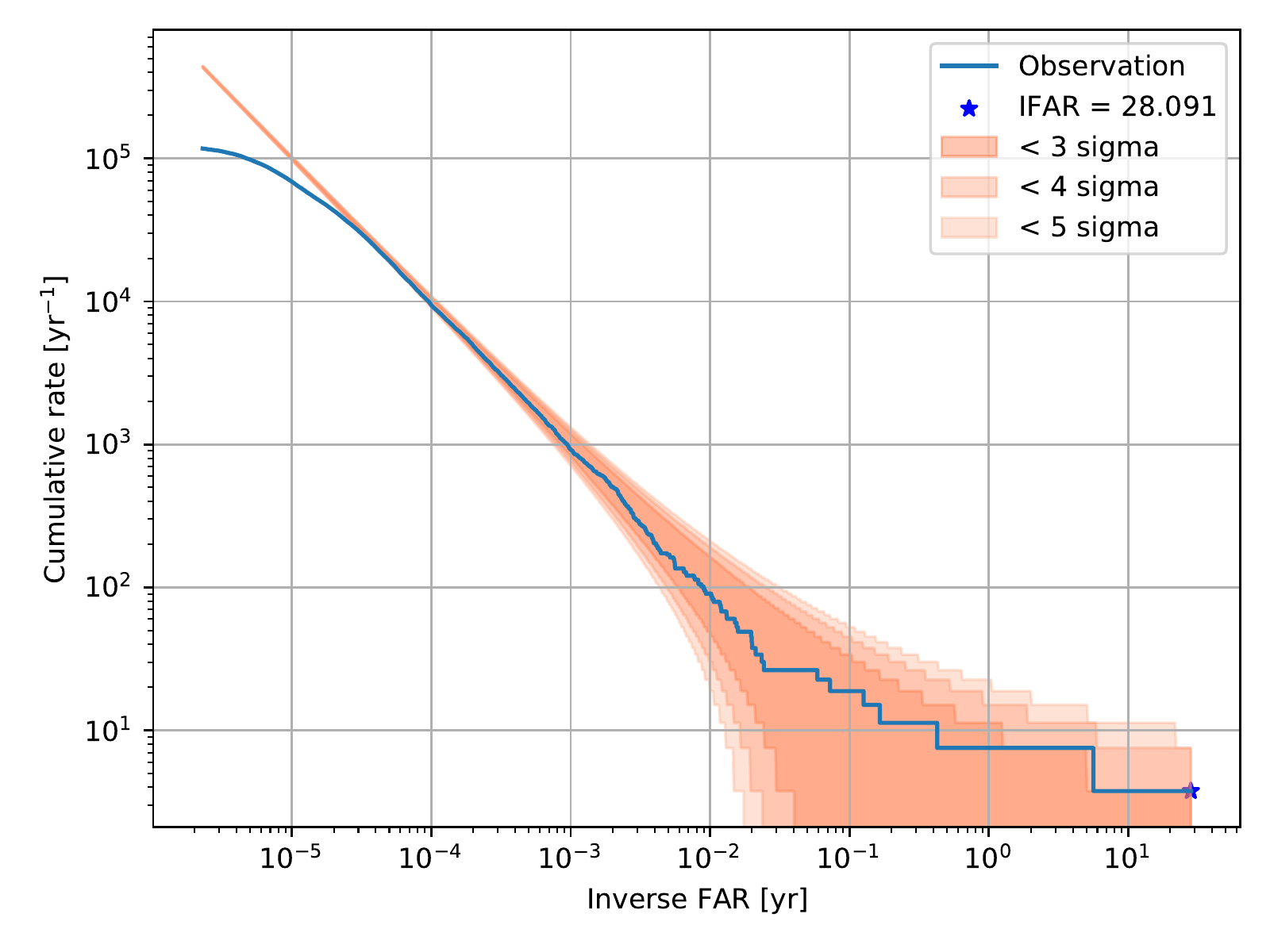}
    \caption{Cumulative rate as a function of the IFAR for foreground (solid line) with configuration 1. The foreground represents associations between Fermi/GBM candidates and LIGO candidates with no time shift.}
    \label{fig:discovery_tuning1}
\end{figure}

The FAR is computed by counting the number of associations in the background with a higher association ranking statistic than the foreground association in question. One can then build the cumulative rate as a function of the IFAR (defined as 1/FAR) for the foreground associations as shown in Figure~\ref{fig:discovery_tuning1}. Here, if we have a joint detection we will see the foreground (blue curve) deviating from the expectation (orange region).

\begin{figure*}
    \centering
    \includegraphics[width=0.45\textwidth]{waterfall_unconstrained_association_0_foreground.pdf}
    \includegraphics[width=0.45\textwidth]{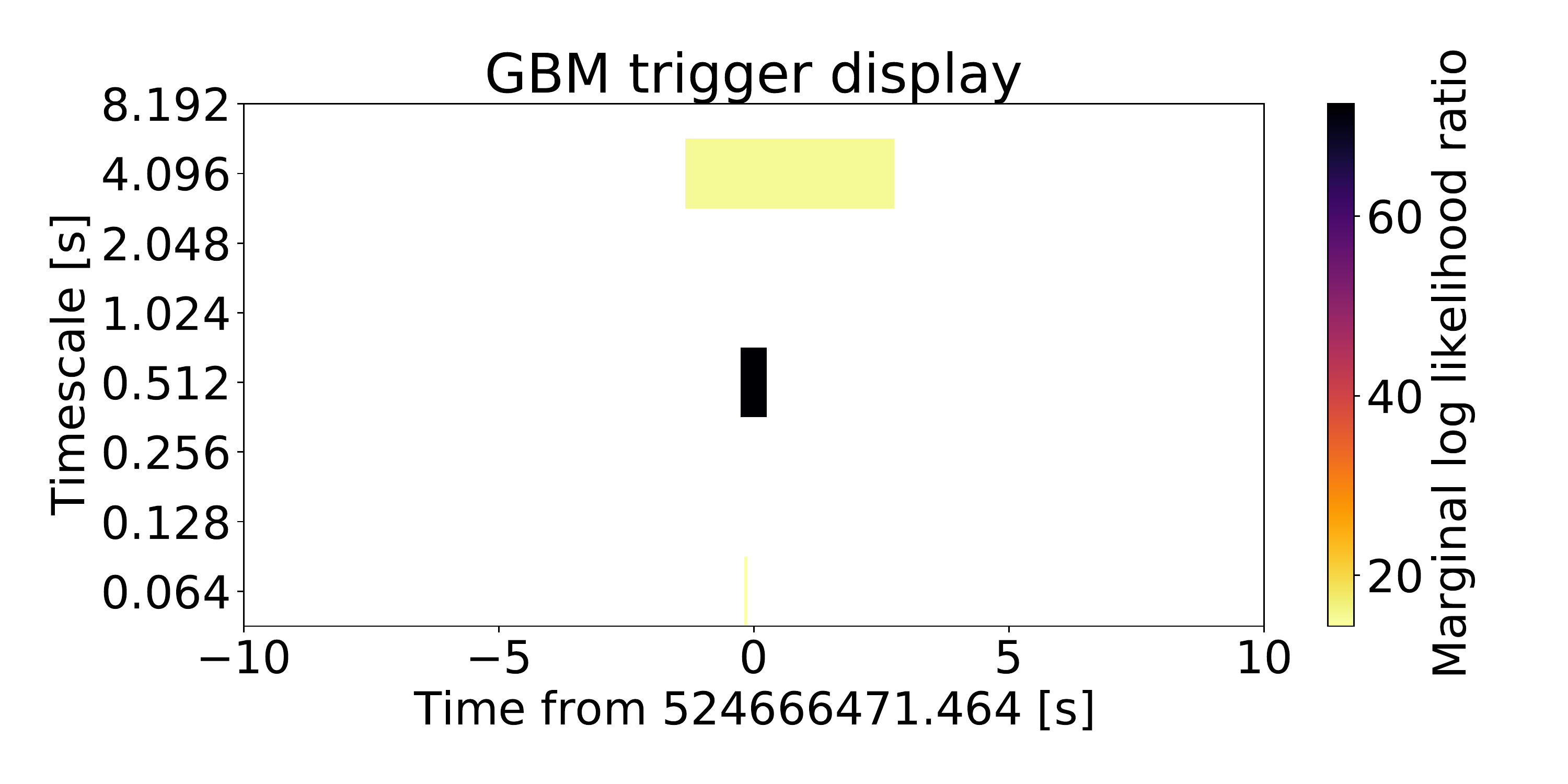}
    \includegraphics[width=0.45\textwidth]{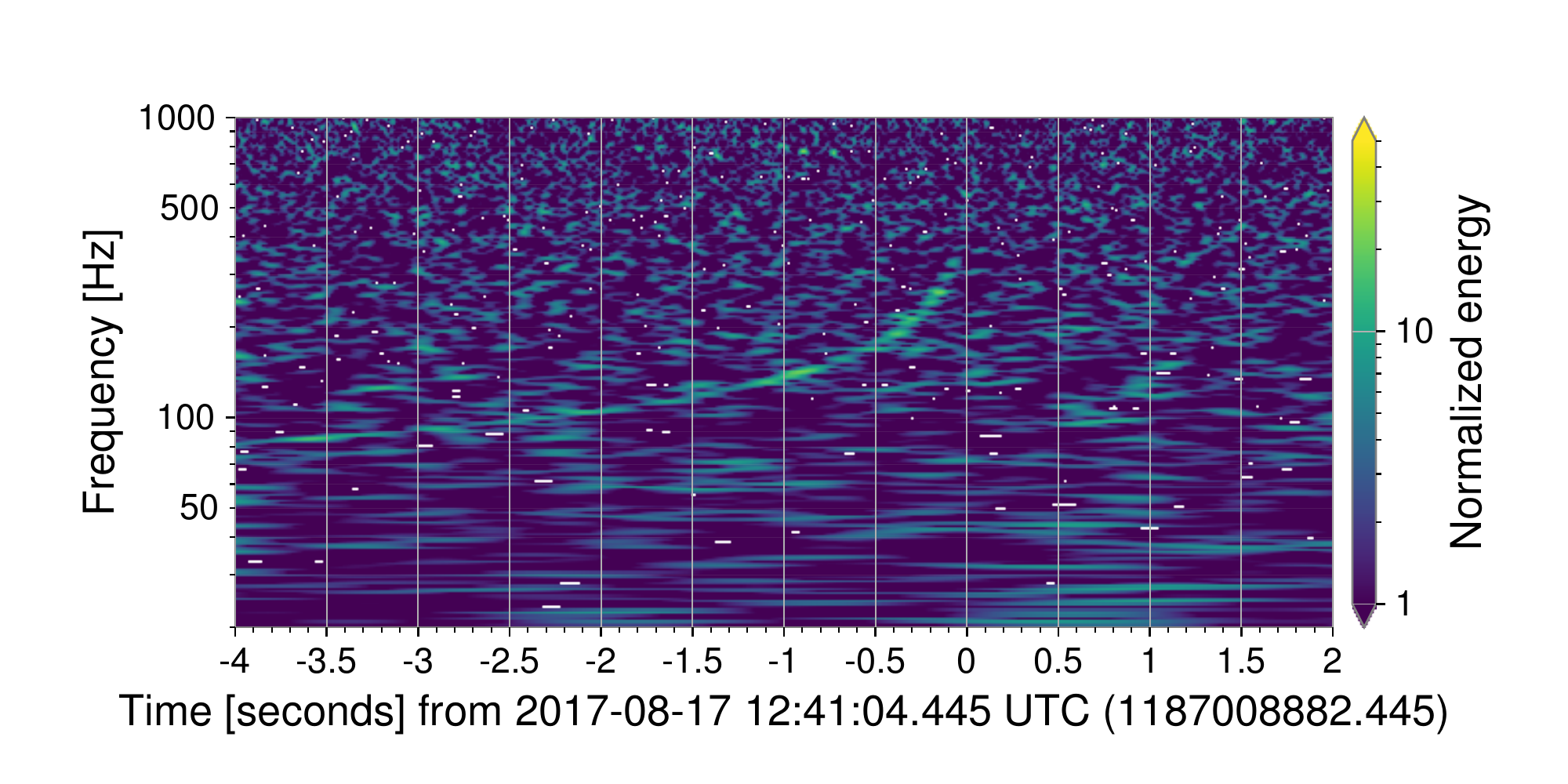}
    \includegraphics[width=0.45\textwidth]{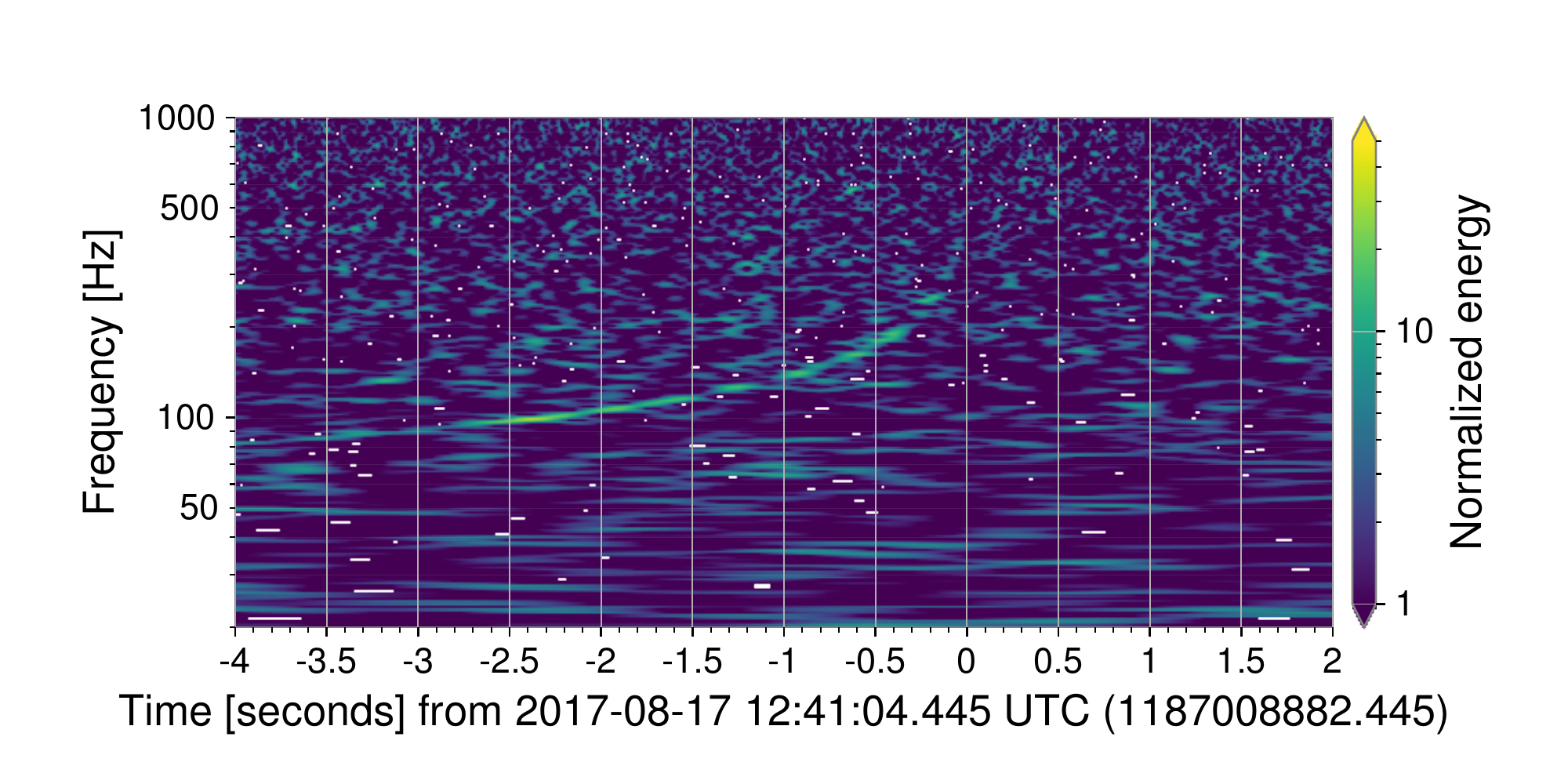}
    \includegraphics[width=0.45\textwidth]{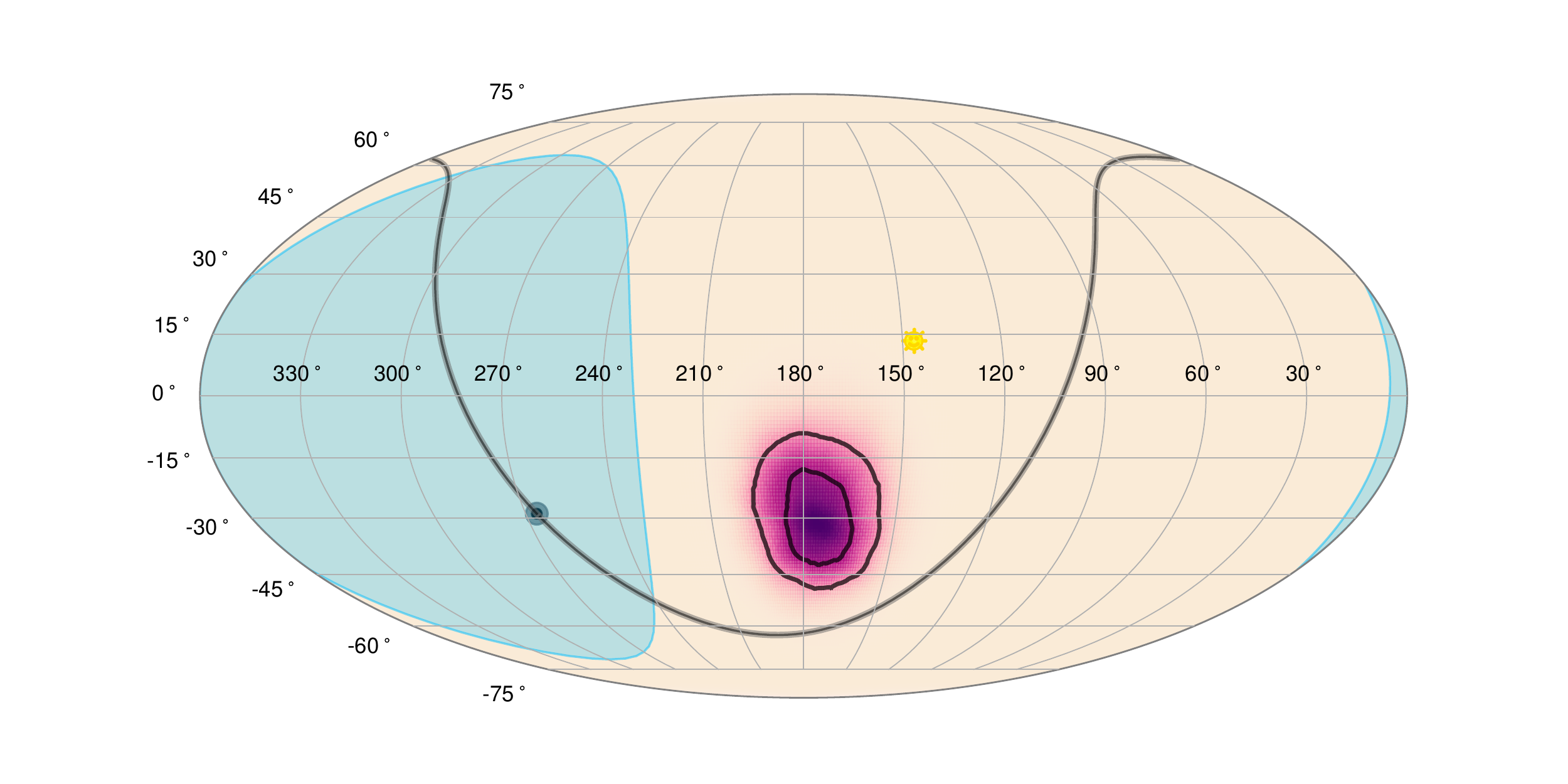}
    \includegraphics[width=0.45\textwidth]{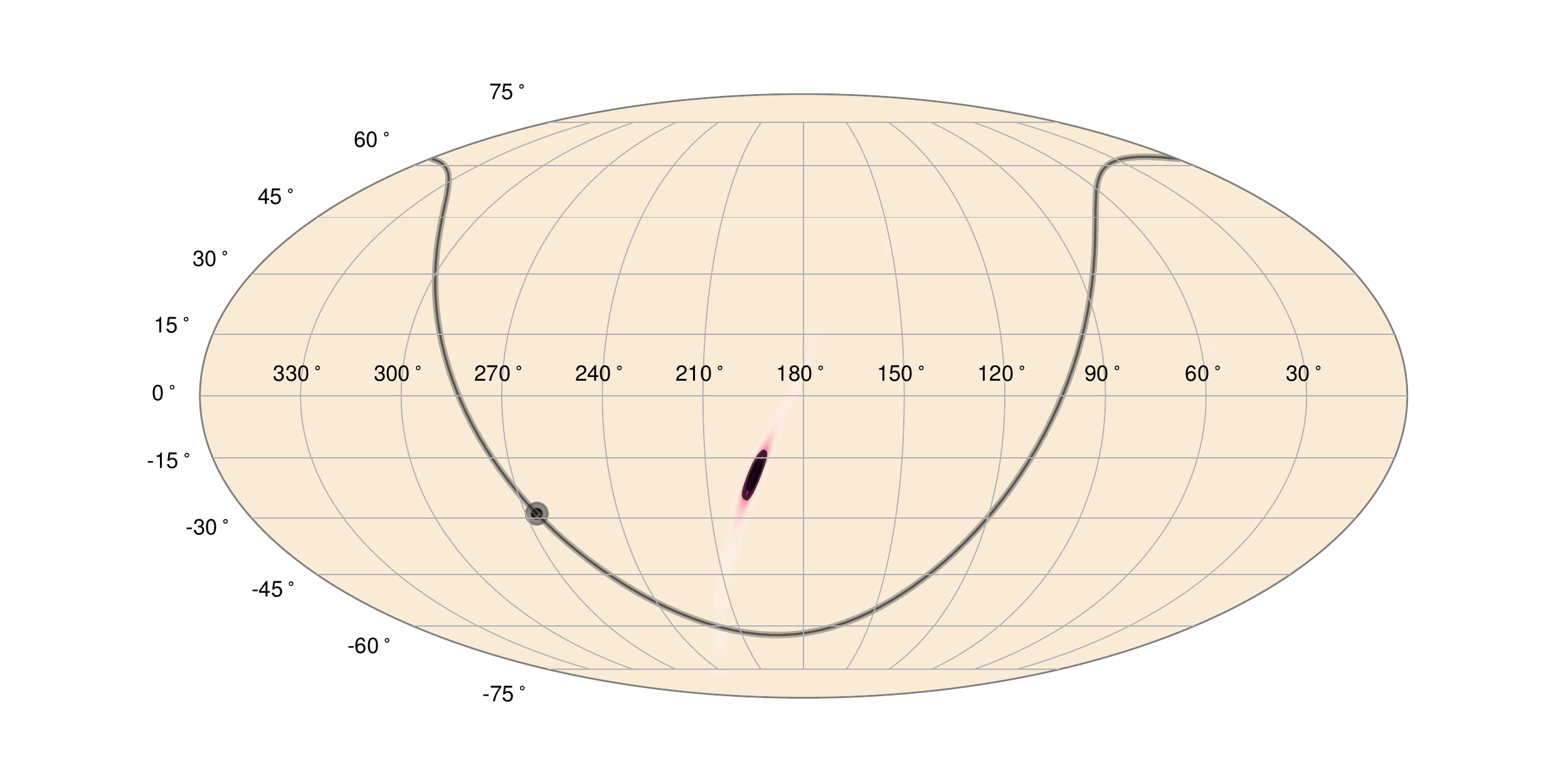}
    \caption{Display of the most significant foreground association with configuration 1 corresponding to GW170817--GRB 170817A. Top: GBM trigger display with waterfall plots of the GBM trigger (left) and clustered GBM trigger (right). Middle: GW trigger display of spectrograms from H1 (left) and L1 (right). Bottom: Sky localization of the association of the GBM trigger (left) and GW trigger (right). The gray solid line represents the Galactic plane, and the Sun is represented by the yellow star in the GBM skymap. The blue region is the Earth's location. The two contour levels represent the 90\% and 50\% credible regions. The darker the purple, the higher the probability.}
    \label{fig:GW170817/GRB170817A_display}
\end{figure*}

The significance of the foreground presented in Figure~\ref{fig:discovery_tuning1} has been computed by separating the associations by pairs of GBM spectral value and GBM duration. In Figure~\ref{fig:discovery_tuning1}, one association deviates at $3\sigma$ from the background with an IFAR $\sim$ 28 yr. Table~\ref{table:tuning1_foreground} presents the first four most significant foreground associations with this configuration. Here, the top two are related to GW170817--GRB 170817A. The second association corresponds to the so-called "soft tail" of GRB 170817A \citep{Abbott_2017, LIGOScientific:2017zic, GRB170817A_1}. The display of GW170817--GRB 170817A is shown in Figure~\ref{fig:GW170817/GRB170817A_display}.
With this configuration, GW170817--GRB 170817A is discovered at barely $3\sigma$, which is not highly significant. Here, the L1 spectrogram has been displayed using the data after subtraction of the glitch.
In Table~\ref{table:tuning1_foreground}, the next associations after GW170817--GRB 170817A are more likely composed of noise in both GBM and GW channels despite their signal-like $Q_{g}$. Their spectrograms, S/N time series and the O2 confident GW events list have been checked. These associations are not significant since they have a small IFAR, so they are more likely to be accidental coincidence. Finally, the third most significant foreground association in Table~\ref{table:tuning1_foreground} appears to have a negative delay, meaning that the GBM trigger is detected before the GW trigger, so it would be surprising if they have a common origin when we put it in the context of astrophysics, in addition to the low IFAR of this association. We stop at the fourth top foreground association because the rest of the table does not contain association with IFAR large enough to be investigated.
Moreover, the beginning of the foreground curve deviates from the expectation computed with the background. This behavior is understood and explanations are given in Appendix~\ref{appendix:modelling}. 

\begin{table*}
\centering
  \begin{tabular}{|c |c |c |c |c |c |c |c |c |c |c |} 
 \hline
 \multicolumn{1}{|c|}{} & \multicolumn{2}{|c|}{GW Properties} & \multicolumn{4}{|c|}{GBM Properties} & \multicolumn{4}{|c|}{Joint Properties} \\
 \hline
Rank & Merger Time & $Q_g$ & Duration (s) & Spectrum & LLR  & $Q_{\gamma}$ & $\Delta t$ (s) & $I^{EA}_{\Omega}$ & $\Lambda$ & IFAR (yr)\\ [0.5ex] 
 \hline
1 & 1,187,008,882.445 & $6.31\times10^{-6}$ & 0.512 & normal  &  72.51 & $1.91\times10^{-3}$ & 2.02 & $17.2$ & $16.0$ & 28.091\\ 
2 & 1,187,008,882.445 & $6.31\times10^{-6}$ & 4.096 & soft & 15.38 & $8.67\times10^{-1}$ & 2.72 & $27.9$ & $13.6$ & 5.618\\
3 & 1,168,226,845.160 & $3.76\times10^{-2}$ & 4.096 & hard & 16.61 & $5.31\times10^{-1}$ & -6.62 & $3.31$ & $1.63$ & 0.426 \\
4 & 1,185,721,264.338 & $7.29\times10^{-3}$ & 0.064 & soft & 13.05 & $1.75$ & 0.638 & $6.22$ & $2.20$ & 0.165 \\ [1ex] 
 \hline
 \end{tabular}
 \caption{Properties of the first four most significant foreground associations with Configuration 1. The two first associations correspond to GW170817--GRB 170817A and its soft tail. $Q_g$: GW Bayes Factor. $Q_{\gamma}$: GBM Bayes Factor. $\Delta t$: time delay between the GBM and the GW trigger. $I^{EA}_{\Omega}$: sky overlap value. $\Lambda$: association rank value.}
 \label{table:tuning1_foreground}
\end{table*}

\subsection{Bayesian Ranking Statistic Without the Separation of the Associations by GBM Spectral Value and GBM Duration}
\label{remove-spec-dur}

\begin{figure}
    \centering
    \includegraphics[width=\columnwidth]{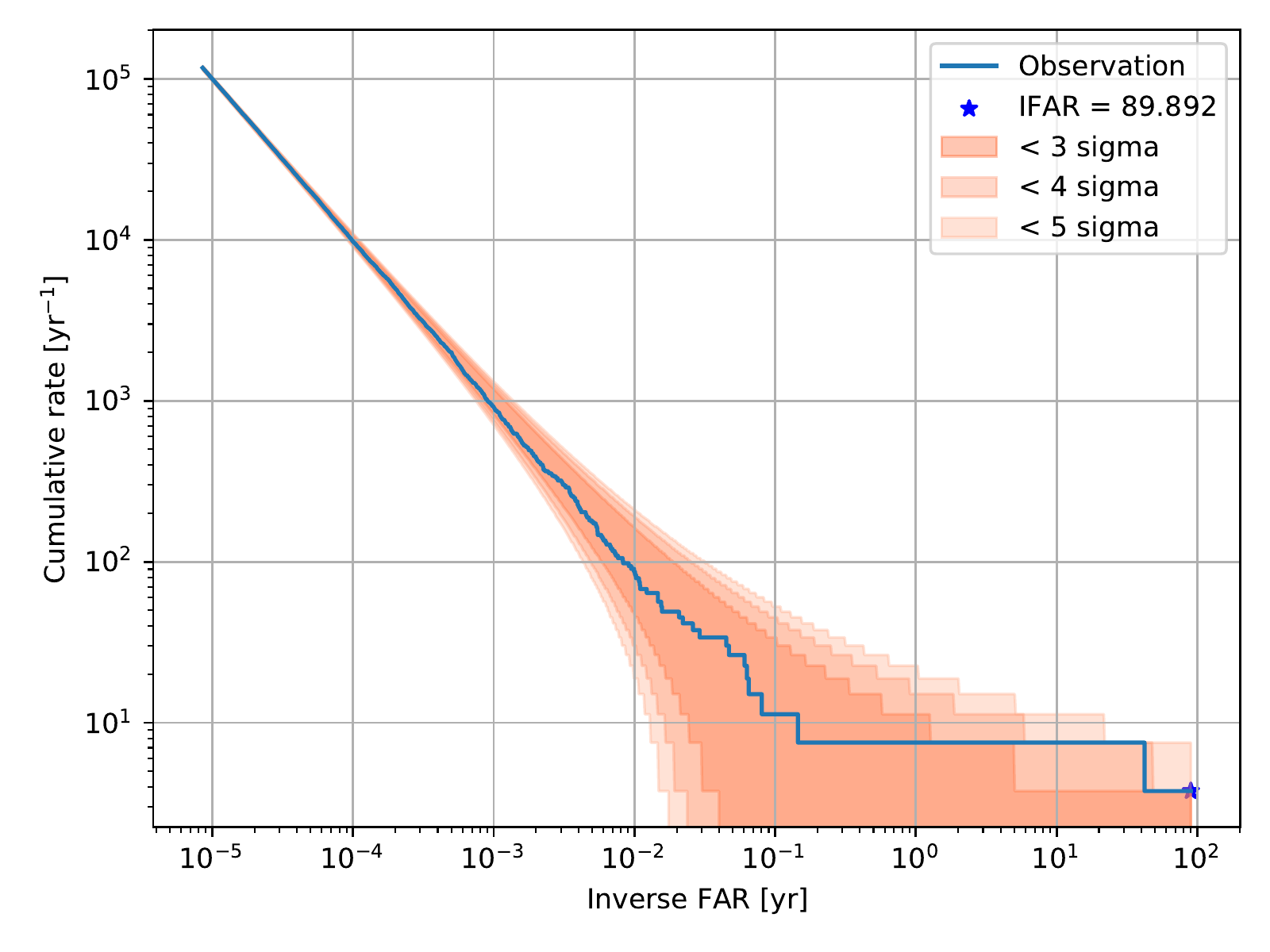}
    \caption{Cumulative rate as a function of the IFAR for the foreground (solid line). The foreground represents associations between Fermi/GBM candidates and LIGO triggers with no time shift. Here the significance is computed with configuration 2 without separating the associations by pairs of GBM spectral value and GBM duration.}
    \label{fig:discovery_tuning1.2}
\end{figure}

When calculating the FAR of an association, we can compare the associations with either the entire background sample or only with only a subset of it and then apply a trial factor based on the number of “bins” into which we split the background sample. In Section~\ref{separation}, we have considered splitting the background by both GBM duration and GBM spectral index. One can also compute the significance of the foreground without separating by GBM spectral value and GBM duration. The results when removing this separation are shown in Figure~\ref{fig:discovery_tuning1.2}. By comparing the joint associations GW170817--GRB 170817A to many more background associations, all GBM spectral values and GBM duration included, we still have a $3\sigma$ deviation from the expected background; however, we increase the IFAR of the joint detection going from 28 yr to 90 yr. Indeed, when we compute the FAR by separating the associations in configuration $n^\circ 1$ and thus treat each search as an independent search, we need to multiply the significance by a trial factor to account for the number of searches. Removing this trial factor increases the value of the IFAR of GW170817--GRB 170817A in configuration $n^\circ 2$. Later we decide not to separate the associations by GBM spectral value and GBM duration.

\subsection{Bayesian Ranking Statistic and Preselection of GW Triggers Based on their FAR}
\label{preselection}

\subsubsection{Limitations of GW170817--GRB 170817A Significance}
The poor significance of GW170817--GRB 170817A presented in Section~\ref{separation} and Section~\ref{remove-spec-dur} demonstrates that our configuration is not as satisfactory as it should be, and investigation of what limits the joint detection's significance shows that some background associations have a higher association ranking statistic than GW170817--GRB 170817A. As an example, Table~\ref{table:tuning1_back_limit} describes a background association with the same GBM duration and spectral value as GRB 170817A and a higher association ranking statistic. The GBM trigger in this association corresponds to a real GRB and, on the GW side, the IFAR of the GW trigger is about $4.265 \times 10^{-5}$yr which is noise-like. 

\begin{table*}
\centering
 \begin{tabular}{|c |c |c |c |c |c |c |c |c |c |} 
 \hline
 \multicolumn{1}{|c|}{} & \multicolumn{2}{|c|}{GW Properties} & \multicolumn{4}{|c|}{GBM Properties} & \multicolumn{3}{|c|}{Joint Properties} \\
 \hline
Rank & Merger Time & $Q_g$ & Duration (s) & Spectrum & LLR  & $Q_{\gamma}$ & $\Delta t$ (s) & $I^{EA}_{\Omega}$ & $\Lambda$\\ [0.5ex] 
 \hline
 1 & 1,176,213,122.254 & $2.98\times10^{-1}$
 & 0.512 & normal  &  176.6 & $1.05\times10^{-3}$
& -6.78 & $17.2$ & $17.8$\\[1ex] 
 \hline
 \end{tabular}
 \caption{Properties of the background association with the same GBM duration and spectral value as GRB 170817A that limits the significance of GW170817--GRB 170817A in configuration $n^{\circ}1$. $Q_g$: GW Bayes Factor. $Q_{\gamma}$: GBM Bayes Factor. $\Delta t$: time delay between the GBM and the GW trigger. $I^{EA}_{\Omega}$: sky overlap value. $\Lambda$: association rank value.}
 \label{table:tuning1_back_limit}
\end{table*}

So in configuration $n^{\circ}1$, when we separate the associations by GBM spectral hardness and duration, such association limits the significance of the joint detection. However, this association is more likely to be an accidental coincidence and should be suppressed by the FAR of its GW candidate. One can conclude that its poor significance is mainly due to the large number of GW triggers we have to consider (mainly composed of noise), and to maximize the significance we decide to apply a cut on the GW triggers based on their FAR value. We choose a threshold of 2 per day, inspired by the choices made in GWTC-3.

\subsubsection{Background Associations}
\begin{figure*}
    \centering
    \includegraphics[width=\textwidth]{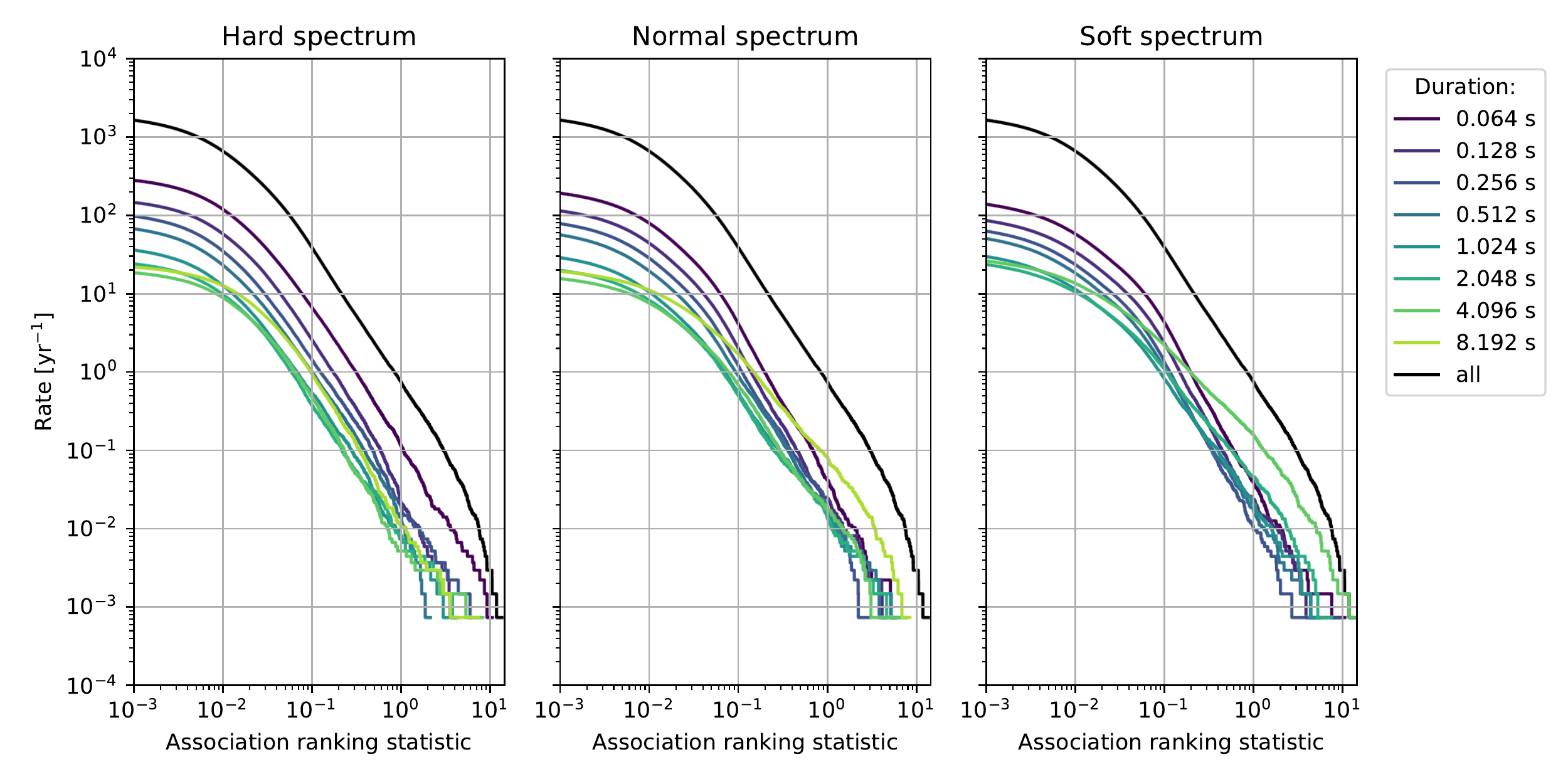}
    \caption{Background association when we cut the GW triggers on their FAR before the analysis (configuration 3). Although this configuration does not separate by GBM spectral hardness and duration we still display the associations separated for illustration. Background rate is shown as a function of the Bayesian ranking statistic. Left: associations with a hard GBM spectrum. Middle: associations with a normal GBM spectrum. Right: associations with a soft GBM spectrum. The black curve represents all the associations regardless of the duration and spectral hardness of their GBM triggers.}
    \label{fig:background_rate_tuning2}
\end{figure*}

Figure~\ref{fig:background_rate_tuning2} shows a lower maximum association ranking statistic than Figure~\ref{fig:background_tuning1}, (from 27.5 in configuration $n^{\circ}$1 to 13.5 in configuration $n^{\circ}$3) with a rate of almost $10^{-3} \mathrm{yr^{-1}}$ and it is dominated by soft 4.096 s and hard 0.064 s GBM triggers. Although this configuration does not separate by GBM spectral hardness and duration we still display the background associations separated for illustration. This justifies our wish to remove this separation since the background associations in the three panels of Figure~\ref{fig:background_rate_tuning2} seem to behave similarly. The background comprises associations with signal-like GBM candidates and noise-like GW candidates. It is composed of very diverse GBM triggers, a large range of LLR and durations (going from 0.064 to 8.192 s) on the GBM trigger side, and all kinds of spectral values (hard, normal, and soft). As previously, a diverse background reassures us that they are more likely accidental coincidences.

\subsubsection{Foreground Associations and Significance}

\begin{figure}
    \centering
    \includegraphics[width = \columnwidth]{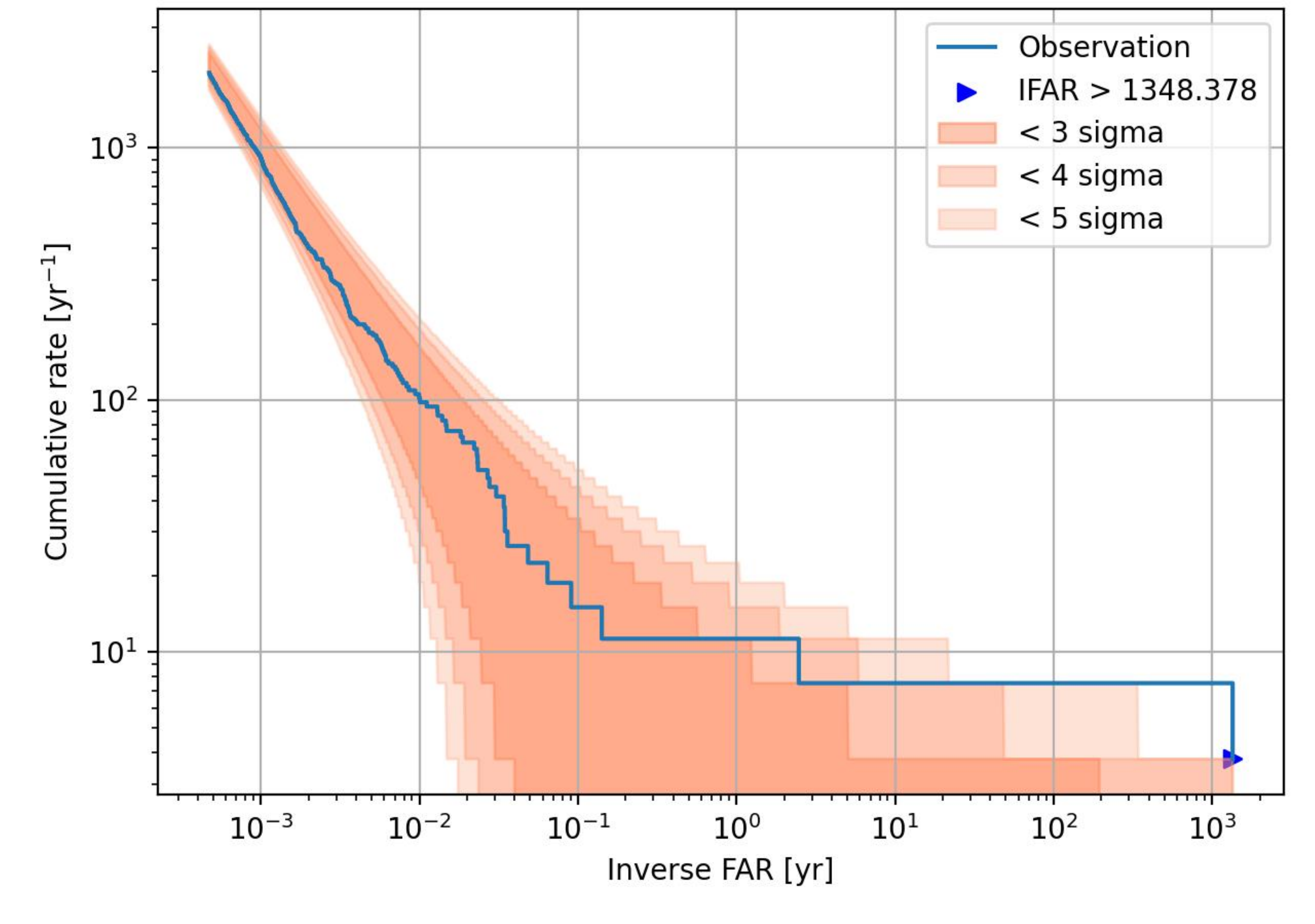}
    \caption{Cumulative rate as a function of the IFAR for the foreground (solid line). The foreground represents associations between Fermi/GBM candidates and LIGO triggers with no time shift. Here preliminary cut is applied to the GW triggers based on their FAR (configuration $n^\circ3$).}
    \label{fig:significance_tuning2}
\end{figure}

Figure~\ref{fig:significance_tuning2} shows that we now have a discovery at more than $4\sigma$ contrary to the results without preselecting the GW triggers. This joint association has an IFAR that is higher than 1348 yr. Since we do not have background associations ranked to compare high enough to compare this foreground association with, this IFAR of the joint detection is now a limit, and the actual IFAR value of this association is greater than this. 

\begin{table*}
\centering
 \begin{tabular}{|c |c |c |c |c |c |c |c |c |c |c |} 
 \hline
 \multicolumn{1}{|c|}{} & \multicolumn{2}{|c|}{GW Properties} & \multicolumn{4}{|c|}{GBM Properties} & \multicolumn{4}{|c|}{Joint Properties} \\
 \hline
Rank & Merger Time & $Q_g$ & Duration (s) & Spectrum & LLR  & $Q_{\gamma}$ & $\Delta t$ (s) & $I^{EA}_{\Omega}$ & $\Lambda$ & IFAR (yr)\\ [0.5ex] 
 \hline
 1 & 1,187,008,882.445 & $6.31\times10^{-6}$
 & 0.512 & normal  &  72.51 & $1.91\times10^{-3}$ & 2.02 & $17.2$ & $16.0$ &  $>$1348\\ 
2 & 1,187,008,882.445 & $6.31\times10^{-6}$ & 4.096 & soft & 15.38 & $8.67\times10^{-1}$ & 2.72 & $27.9$ & $13.6$ & $>$1348\\
3 & 1,187,008,882.445 & $6.31\times10^{-6}$ & 0.064 & hard & 14.32 &  $9.20\times10^{-1}$ & 1.86 & $2.86$ & $1.40$ & 2.474\\
4 & 1,185,284,217.254 & $1.18\times10^{-2}$ & 0.064 & normal & 9.457 & $10.0$ & 12.71 & $5.05$ & $0.261$ & 0.142 \\ [1ex] 
 \hline
 \end{tabular}
 \caption{Properties of the first four most significant foreground associations with configuration $n^\circ3$. The top 3 contain GW170817--GRB 170817A. $Q_g$: GW Bayes Factor. $Q_{\gamma}$: GBM Bayes Factor. $\Delta t$: time delay between the GBM and the GW trigger. $I^{EA}_{\Omega}$: sky overlap value. $\Lambda$: association rank value.}
 \label{table:tuning2_foreground}
\end{table*}

\begin{figure}
    \centering
    \includegraphics[width=\columnwidth]{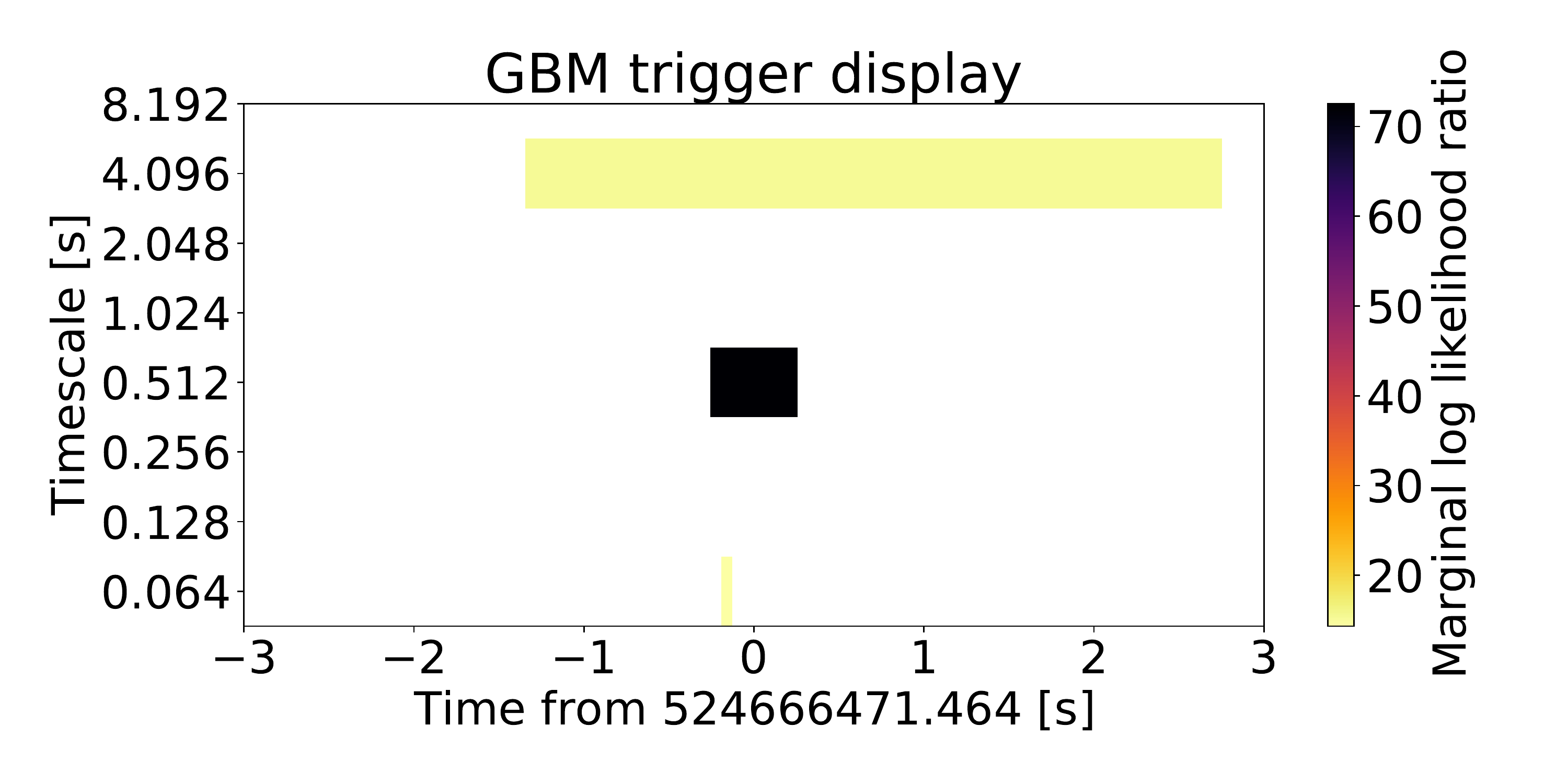}
    \caption{Waterfall plot showing GRB 170817A. The three rectangles at three different timescales are the results of the clustering that produced three GBM triggers for this event. These three are at the top of the table of the most significant foreground associations.}
    \label{fig:waterfall_GRB170817A_3_triggers}
\end{figure}

Table~\ref{table:tuning2_foreground} shows the first four most significant foreground associations. Now the top three are related to GW170817--GRB 170817A. The cut of the GW trigger has brought a third weak extra GBM trigger connected to GRB 170817A and coupled with GW170817. Figure~\ref{fig:waterfall_GRB170817A_3_triggers} shows the three GBM triggers of GRB 170817A that correspond to the three most significant associations. The fourth ranked foreground association does not contain significant candidates in both channels. Its $Q_{\gamma}$ is noise-like and its $Q_{g}$ is signal-like but L1 and H1 data do not show any excess of power at the time of the GW trigger. Finally, its joint IFAR is also not extremely high, so this association is more likely to be accidental.

Discarding GW triggers highly likely to be noise before running this analysis allows us to remove loud background (time-shifted) associations containing noise in the GW channel and limiting the significance of GW170817--GRB 170817A such as those in Table~\ref{table:tuning1_back_limit}. We finally found an optimal configuration that enables the maximization of GW170817--GRB 170817A's significance. Table~\ref{table:tuning2_foreground} also demonstrates that GW170817--GRB 170817A was the only GW--GRB astrophysical association that has been observed in the GW and GBM data during O2 because the fourth most significant foreground association does not have a sufficiently high IFAR to be investigated as an interesting association.

\section{Conclusion}
\label{conclusion}
We presented a method to search for associations in a symmetric way between GW triggers from LIGO and Virgo interferometers and data from Fermi/GBM. We sorted the associations thanks to a ranking statistic introduced in \citep{Ranking} based on the time and spatial overlap and the significance of the GW and the GBM candidates. The method described here has been applied to PyCBC and Fermi/GBM triggers from O2 to check its validity against GW170817--GRB 170817A. Several configurations have been tested to maximize the significance of this joint detection. Indeed, when we have to analyze a large amount of noise on the GW side, GW170817--GRB 170817A is not highly significant. When we preselect the GW triggers on their FAR to reduce their rate, we rediscovered GW170817--GRB 170817A with the three most significant foreground associations related to this joint detection. The first two have an IFAR $>$ 1348 yr. Other configurations could be tested to maximize the significance of the joint detection such as removing the GW triggers with masses incompatible with neutron stars or using a more realistic prior for the time offset, since currently we allow the GBM trigger to happen long before the GW trigger. 
In \citep{Zhang_2019}, the time delay between the GW and the GRB is written as a function of the time for the central engine to launch a relativistic jet $\Delta t_{\mathrm{jet}}$, the time for the jet to penetrate through and break out from the surrounding medium $\Delta t_{\mathrm{bo}}$, and the time after breakout for the jet to reach the energy dissipation radius where the observed $\gamma$-rays are emitted $\Delta t_{\mathrm{GRB}}$, thanks to the following formula: $\Delta t_{\mathrm{GW-GRB}} = (\Delta t_{\mathrm{jet}} + \Delta t_{\mathrm{bo}} + \Delta t_{\mathrm{GRB}})(1+z)$. Starting from this review \citep{Zhang_2019} and understanding the jet and GRB physics in BNS systems would lead to an estimation of $\Delta t_{\mathrm{GW-GRB}}$ that could be used as our time offset prior in future analysis.

We are also considering improving the calculation of the GW Bayes factor. Thus, a new calculation of $Q_{g}$ with a KDE is proposed in Appendix~\ref{appendix:QG} and we may test it in future work. Finally, we did a search that is optimized for short GRBs since they are believed to be produced by neutron star mergers, contrary to long GRBs that are due to CCSNe. However, this classification is now questioned since \citep{Rastinejad2022} recently observed cases where a kilonova, the remnant of neutron star mergers, might has been associated with a long GRB. One other improvement would be to extend the search to all types of GRBs by, for example, taking a sample of short and long GRBs as a signal training sample in the calculation of GBM Bayes factor. 

Furthermore, improvements in the filtering and clustering algorithm of the GBM Targeted Search can be explored. Indeed, Figure~\ref{fig:waterfall_GRB170817A_3_triggers} shows that the clustering generated three triggers instead of one for the same GRB. This indicates that some residual correlations are present in the GBM triggers. 

The method presented with the optimal configuration will be used on the GWTC-3 released PyCBC, GstLAL, and MBTA triggers and Fermi/GBM data covering the duration of O3, in the hope of discovering associations that have not been discovered yet.
Finally, the approach we presented in this paper is not limited to Fermi/GBM data. Other missions like the future SVOM mission \citep{2022IJMPD..3130008A} can use a similar approach if they produce their own form of GBM Bayes factor $Q_{\gamma}$ and provide independent sky localizations. Thus, another potential improvement would be considering information from multiple GRB monitors.

\begin{acknowledgements}
T.D. thanks Francesco Pannarale at the Sapienza University of Rome for the kind hospitality during part of this work.
We thank Rosa Poggiani and Fiona Panther for their comments on the analysis and text. We also thank Barbara Patricelli for her useful comments. The authors are grateful for computational resources provided by the LIGO Laboratory and supported by National Science Foundation grants PHY-0757058 and
PHY-0823459. 
R.H. acknowledges funding from the European Union's Horizon 2020 research and innovation program under the Marie Skodowska-Curie grant agreement No. 945298-ParisRegionFP. \

This paper has LIGO document number LIGO-P2300018 and a Virgo document number VIR-0453B-23.
\end{acknowledgements}

\appendix

\section{KDE-based Method to Compute $Q_g$}
\label{appendix:QG}
The current GW Bayes factor does not allow us to discriminate correctly between noise and an astrophysical GW signal. For instance, Table~\ref{table:tuning1_background} shows that the four highest ranked background associations have signal-like GW Bayes factor ($Q_g < 1$). Thus, a new way to compute $Q_g$ should be investigated for future analysis to properly separate the noise-like GW triggers from the interesting GW candidates.

The skymaps produced by \texttt{Bayestar} \citep{Bayestar} provide the BCI (the current $Q_g$) along with the BSN. Consequently, it should be possible to use a KDE-based method to compute $Q_G$ with a KDE trained in the $\mathrm{log}_{10}(\mathrm{BCI})-\mathrm{log}_{10}(\mathrm{BSN})$ plane. We also preprocess the data using the logarithm here in order to avoid dealing with values very different from each other. The expected behavior is summarized in the following list:

\begin{itemize}
    \item Gaussian noise should have both small $\mathrm{log}_{10}(\mathrm{BCI})$ and $\mathrm{log}_{10}(\mathrm{BSN})$ values.
    \item Glitches should have a small $\mathrm{log}_{10}(\mathrm{BCI})$ value and a high $\mathrm{log}_{10}(\mathrm{BSN})$ value.
    \item Astrophysical GW signals should in principle have both high $\mathrm{log}_{10}(\mathrm{BCI})$ and high $\mathrm{log}_{10}(\mathrm{BSN})$ values.
\end{itemize}

Thus, we constructed a GW Bayes factor using these parameters. We built a signal sample with 1000 BNS injections and GW events coming from all runs. The background sample is composed of 2000 O2 GW triggers with a FAR $>$ 2/day. The injections of the signal sample were done in Gaussian and stationary noise with the parameters summarized in Table~\ref{tab:injections}. A cut on the optimal S/N has been applied: only the signals with a network S/N above 8 and an S/N above 5.5 in at least one of the interferometers were kept. The remaining signals were localised with \texttt{Bayestar} to generate the skymaps.

\begin{table}[h]
\centering
\begin{tabular}{cc}
Parameter      & Distribution   \\
\hline
\hline
Polarization  & $\mathcal{U}([0, 2\pi])$ \\
Phase at coalescence  & $\mathcal{U}([0, 2\pi])$ \\
$\cos{\iota}$  & $\mathcal{U}([-1, 1])$ \\
Position  & $\mathcal{U}(\mathbb{S}(0, [40, 250]$Mpc))  \\
Mass  & $\mathcal{N}(\mu=1.4$$M_{\odot},\,\sigma=0.1$$M_{\odot})$ \\
Spin  & $\mathcal{N}(\mu=0,\,\sigma=0.01)$ \\
\hline
\end{tabular}
\caption{Injection parameters for the positive sample used to train the KDE. The $\iota$ parameter corresponds to the inclination of the binary. The position of the simulated signals are generated uniformly in a shell $\mathbb{S}$ with inner radius of 40Mpc and outer radius 250Mpc.}
\label{tab:injections}
\end{table}

\begin{figure*}
    \centering
   \includegraphics[width=0.45\textwidth]{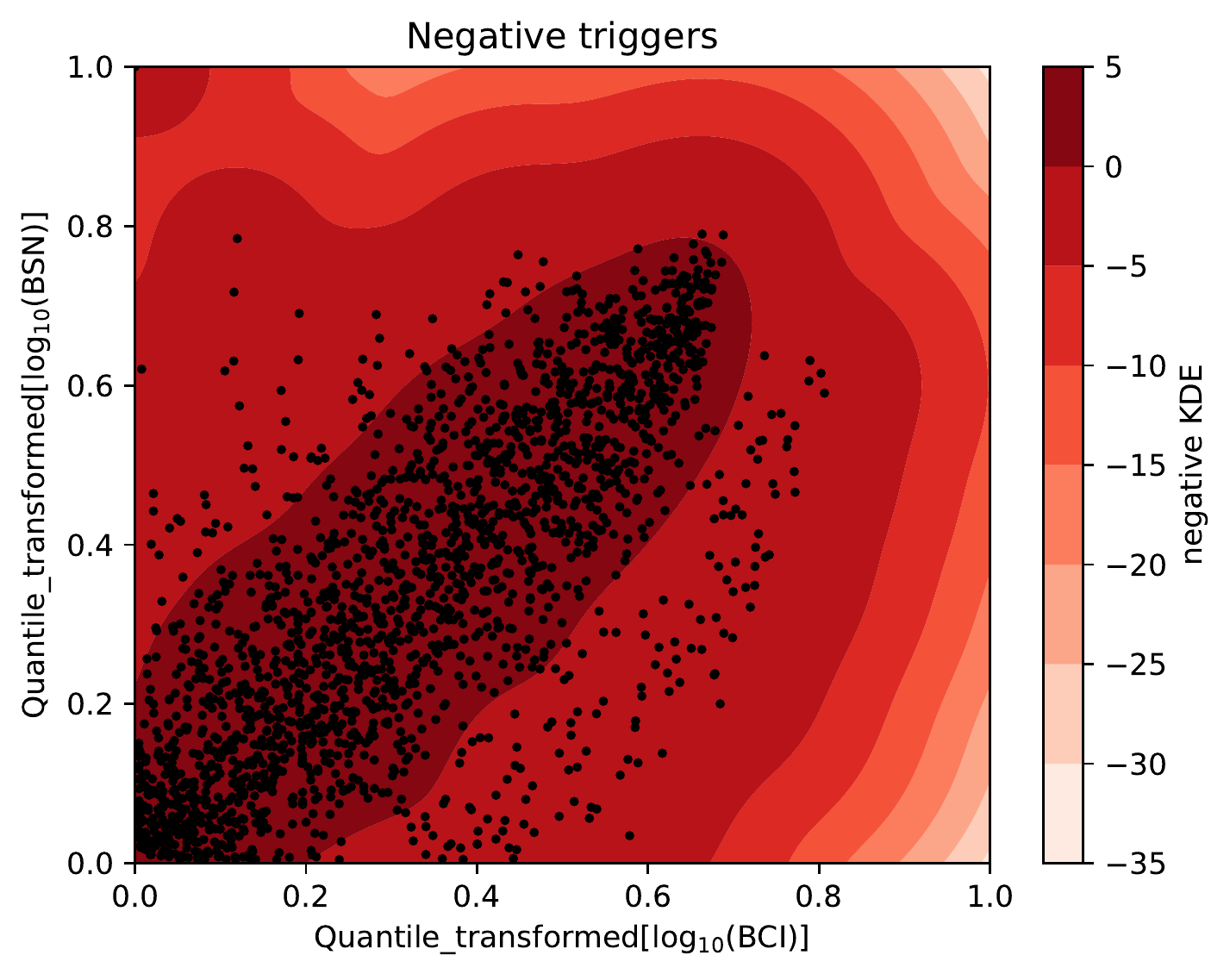}
    \includegraphics[width=0.45\textwidth]{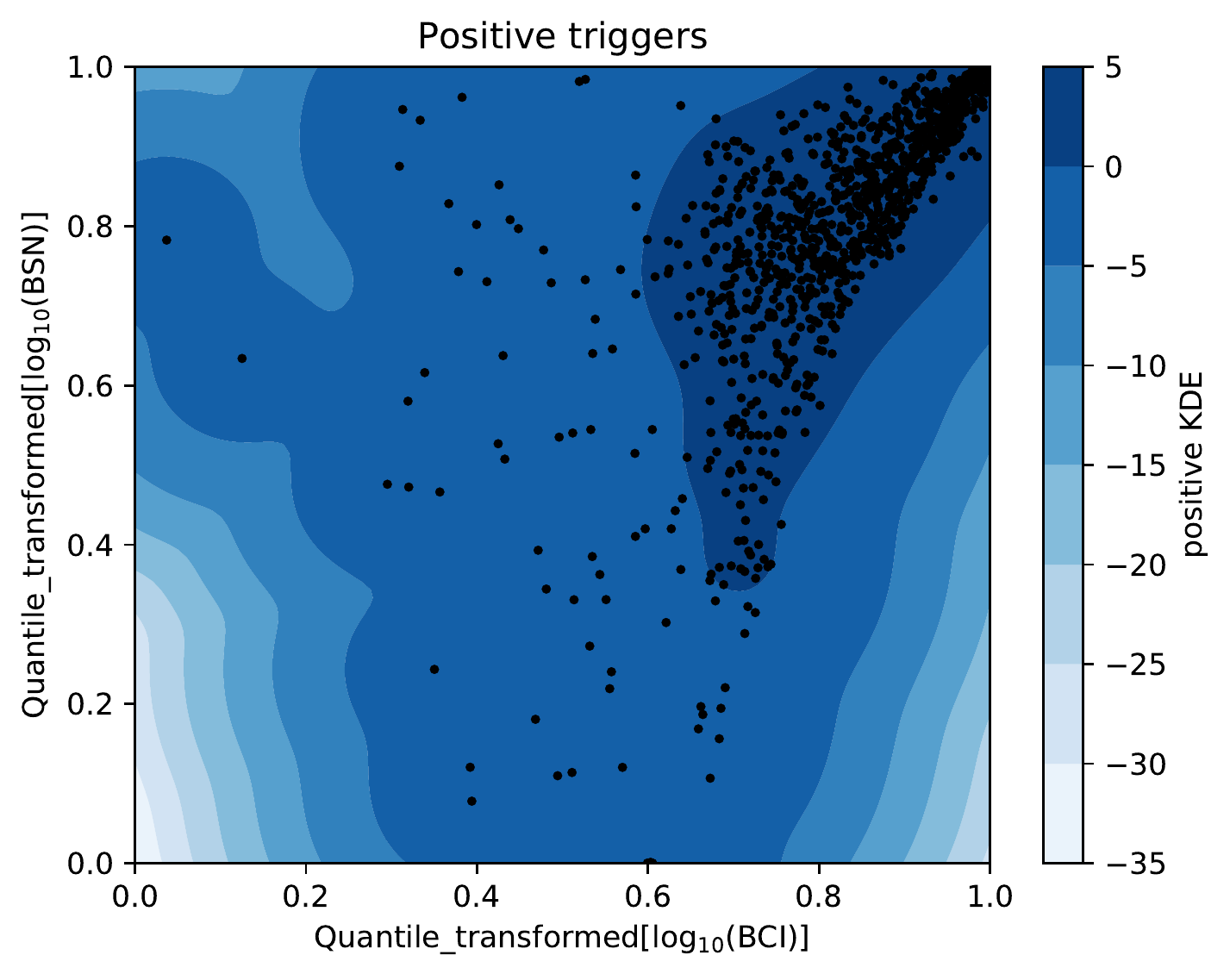}
    \includegraphics[width=0.45\textwidth]{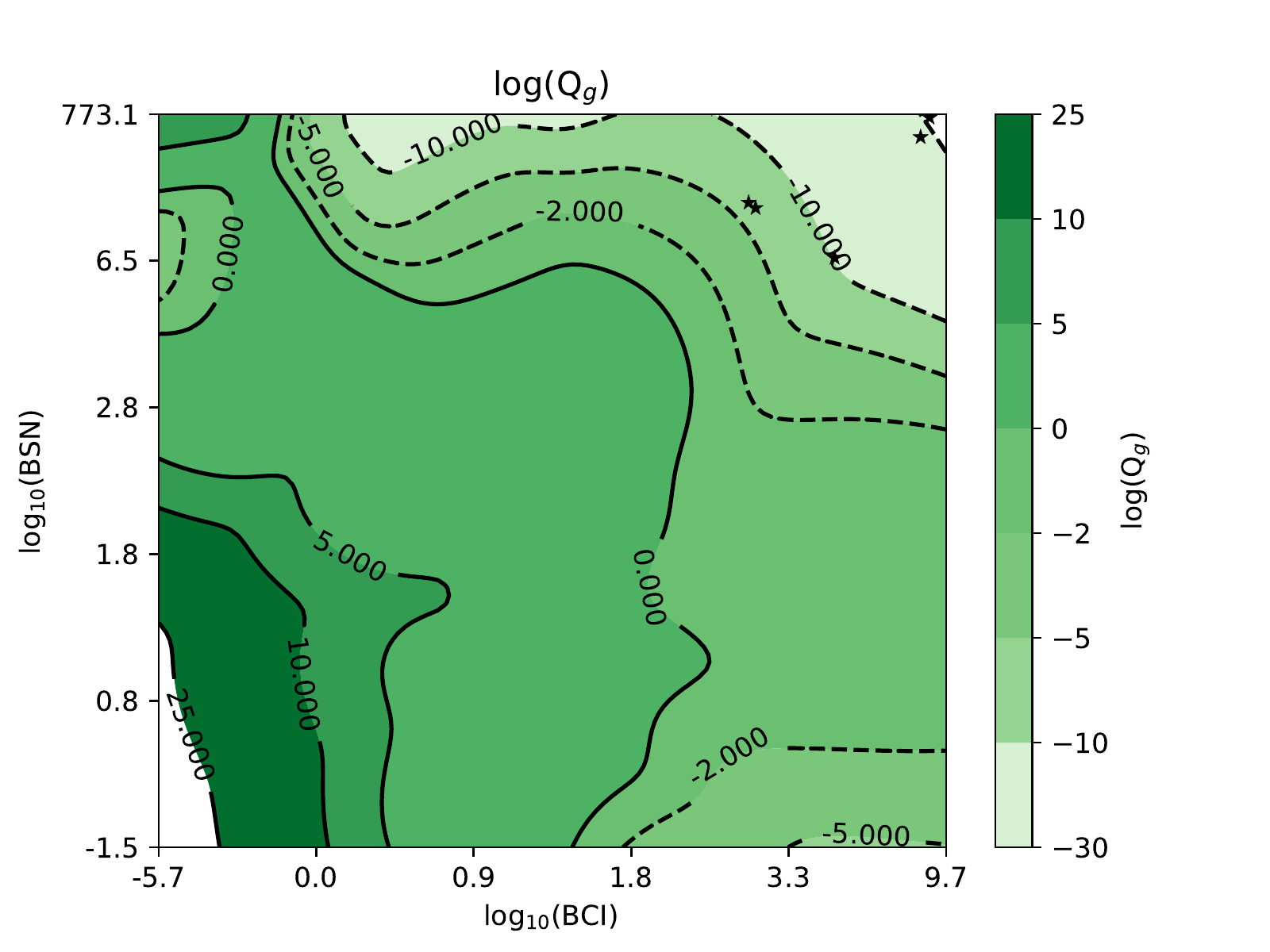} 
    \caption{Summary of the KDE training and evaluation on a simulated sample. The signal-like region corresponds to a $Q_g$ close to 0 (i.e, log($Q_g$) negative). Top left: on the negative sample: O2 GW triggers with a FAR $>$ 2/day. Top right: on the positive sample: O3 injections and confident GW events from O2 and O3. Bottom: Bayes factor distribution in the $\mathrm{log_{10}(BCI)}-\mathrm{log_{10}(BSN)}$ plane. Here the stars are a validation sample composed of four GW events that were not included in the positive sample.}
    \label{fig:O3GWBFfg}
\end{figure*}

The results of the GW Bayes factor are presented in Figure~\ref{fig:O3GWBFfg}. Here we preprocessed the data with a quantile transform \citep{scikit-learn} that flattens the distributions and moves them between 0 and 1. We can clearly see a separation between the signal-like and the noise-like regions. We also checked that this Bayes factor behaves as expected by applying the KDE on a validation sample (stars in the bottom plot of Figure~\ref{fig:O3GWBFfg}). Here, all the GW events from the validation sample are in the signal-like region with strongly negative $\mathrm{log}_{10}(Q_g)$, which is consistent with the expectations. However, the ln$(Q_g) = 0$ curve in Figure~\ref{fig:O3GWBFfg} is almost vertical from log$_{10}$(BSN) $\approx$ -1.5 to log$_{10}$(BSN) $\approx$ 6.5, so except for very high BSN, $Q_g$ seems to depend only on the BCI. As mentioned in Section~\ref{GWBF}, computing a KDE is complicated because some assumptions needs to be made. We need to generate skymaps for GW injections, noise, and glitches. We also need to choose a \emph{bandwidth} for the KDE computation, and results can be strongly bandwidth-dependent. Thus, using only the BCI might be an acceptable approximation and reduces both the number of assumptions we need to make and the computational cost of $Q_g$. 

Another method that could be investigated in the future consists in using the calculation pf the probability of astrophysical origin, \emph{p-astro} \citep{PhysRevX.9.031040}, which is based on binary system rate densities. However, for now the estimation of the BNS merger rate is not robust enough because we have not observed enough of these events, but hopefully we will be able to improve this estimation through future GW observing runs. 

\section{Discrete Computation of Overlaps of HEALPix Skymaps}
\label{appendix:skymap}
A discrete HEALPix localization array from \texttt{Bayestar}  or Fermi/GBM,
hpa, is normalized such that each pixel contains the probability for the source to be inside that pixel, which means that : 
\begin{equation}
    \sum_{i}^{N}\texttt{hpa}[i] = 1
    \label{Discrete1}
\end{equation}
where $N$ is the number of pixels in the HEALPix array. We must compare this
expression with the normalization of a continuous probability density,
\begin{equation}
    \int P(\Omega \mid \texttt{hpa})d\Omega = 1.
    \label{Discrete2}
\end{equation}
Hence, to go from the continuous version to the discrete one, we need
the following replacement:
\begin{equation}
    \int \rightarrow \sum_{i}^{N}
    \label{Discrete3}
\end{equation}
\begin{equation}
    P(\Omega\mid \texttt{hpa})d\Omega \rightarrow \texttt{hpa}[i]
    \label{Discrete4}
\end{equation}
The discrete form of Equation~\ref{Iomega} is then, assuming we have three HEALPix arrays for the prior and posteriors,
\begin{equation}
    I_{\Omega} = \sum_{i}^{N}\frac{\texttt{hpa}\_\texttt{gw}[i]\texttt{hpa}\_\texttt{gbm}[i]}{\texttt{hpa}\_\texttt{prior}[i]}.
    \label{eqn:Discrete5}
\end{equation}
With a uniform prior on the sky, Equation~\ref{Iomega} becomes 
\begin{equation}
    I_{\Omega}^{AS} = 4\pi\int P(\Omega\mid \mathrm{D}_g)P(\Omega\mid \mathrm{D}_\gamma)d\Omega
    \label{IOmegaAS}
\end{equation}
In the case of Equation~\ref{IOmegaAS}, the $d\Omega$’s no longer cancel out: one remains at the denominator
and needs to be replaced with the HEALPix pixel area $\Delta\Omega$. Therefore
\begin{equation}
    I_{\Omega}^{AS} = \frac{4\pi}{\Delta \Omega}\sum_{i}^{N}\texttt{hpa}\_\texttt{gw}[i]\texttt{hpa}\_\texttt{gbm}[i]
    \label{Descrete5}
\end{equation}
The case of Equation~\ref{IOmegaEA} is similar, but we can note that the sum must avoid the pixels covered
by the Earth:
\begin{equation}
    I_{\Omega}^{EA} = \frac{4\pi}{\Delta \Omega}\sum_{i\notin\Earth }^{N}\texttt{hpa}\_\texttt{gw}[i]\texttt{hpa}\_\texttt{gbm}[i]
    \label{Discrete6}
\end{equation}

\section{Modeling the Curve of the Cumulative Rate as a Function of the Inverse False Alarm Rate}
\label{appendix:modelling}
\begin{figure}
    \centering
    \includegraphics[width=0.6\textwidth]{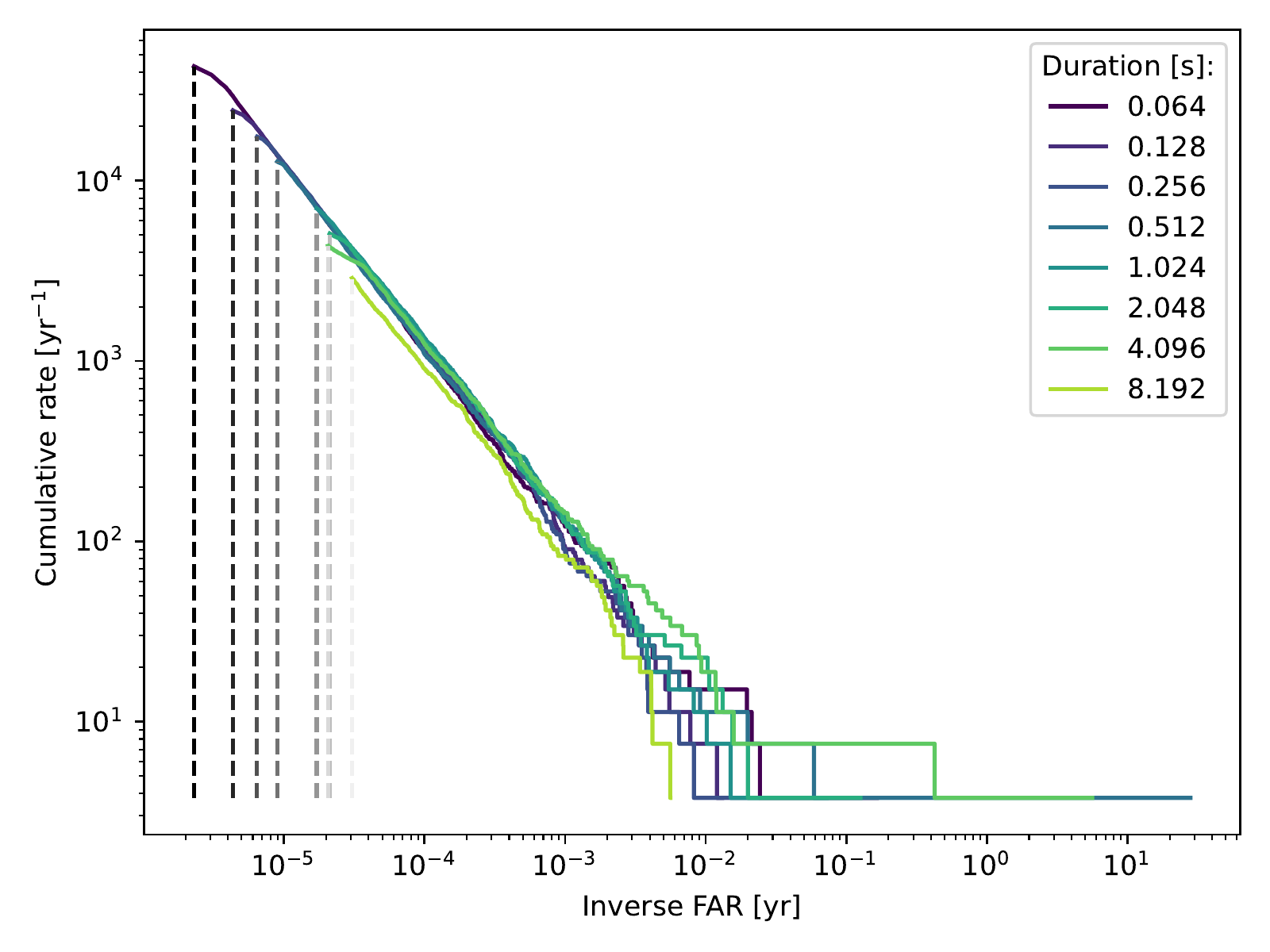}
    \caption{Cumulative rate as a function of IFAR [yr] for different GBM durations. The gray dashed lines show how to bin the curve depending on the duration of the GBM triggers.}
    \label{fig:rates}
\end{figure}

As mentioned in Section~\ref{separation}, Figure~\ref{fig:background_tuning1} shows a rate turnover on the left of the plot. 
This is due to the fact that each pair (GBM temporal-spectral value) does not have the same rate. GRBs with short time-scale, for instance, are occurring more often due to their short duration. This behavior is explicit when looking at Figure~\ref{fig:background_tuning1} in which we can see that each curve does not start at the same rate. So when we separate the associations by GBM spectral value and GBM duration each pair does not start to contribute to the cumulative rate at the same IFAR. Investigations are made to model the expectation curve (in orange) to have the same turnover. We consider here the results from configuration $n^{\circ}$1 (i.e. no cut on GW triggers, before applying the search and separation between GBM spectral value and duration). Figure~\ref{fig:rates} shows that when we sort by GBM duration, all the curves do not start at the same rate. One can create bins with edges defined by the dashed lines in Figure~\ref{fig:rates}. Each bin is represented by its mean IFAR value. Then, one can weigh the bins by the cumulative number of triggers that contribute to each bin.

This can be summarized by the Eqs.~\ref{rate1} and \ref{rate2}, where $i$ represents each bin's index of the lower edge:
\begin{equation}
    R_{i} = \mathrm{Rate}(\frac{\texttt{edge}[i]+\texttt{edge}[i+1]}{2})
    \label{rate1}
\end{equation}
\begin{equation}
    R_{i'} = R_{i} \times \frac{\sum_{0}^{i+1} \texttt{nb}\_\texttt{triggers}}{\sum_{0}^{\infty}\texttt{nb}\_\texttt{triggers}}
    \label{rate2}
\end{equation}
Here we have represented each bin by the IFAR value in its center, so we take only one value per bin to model the left of the curve. 

\begin{figure}
    \centering \includegraphics[width=0.6\textwidth]{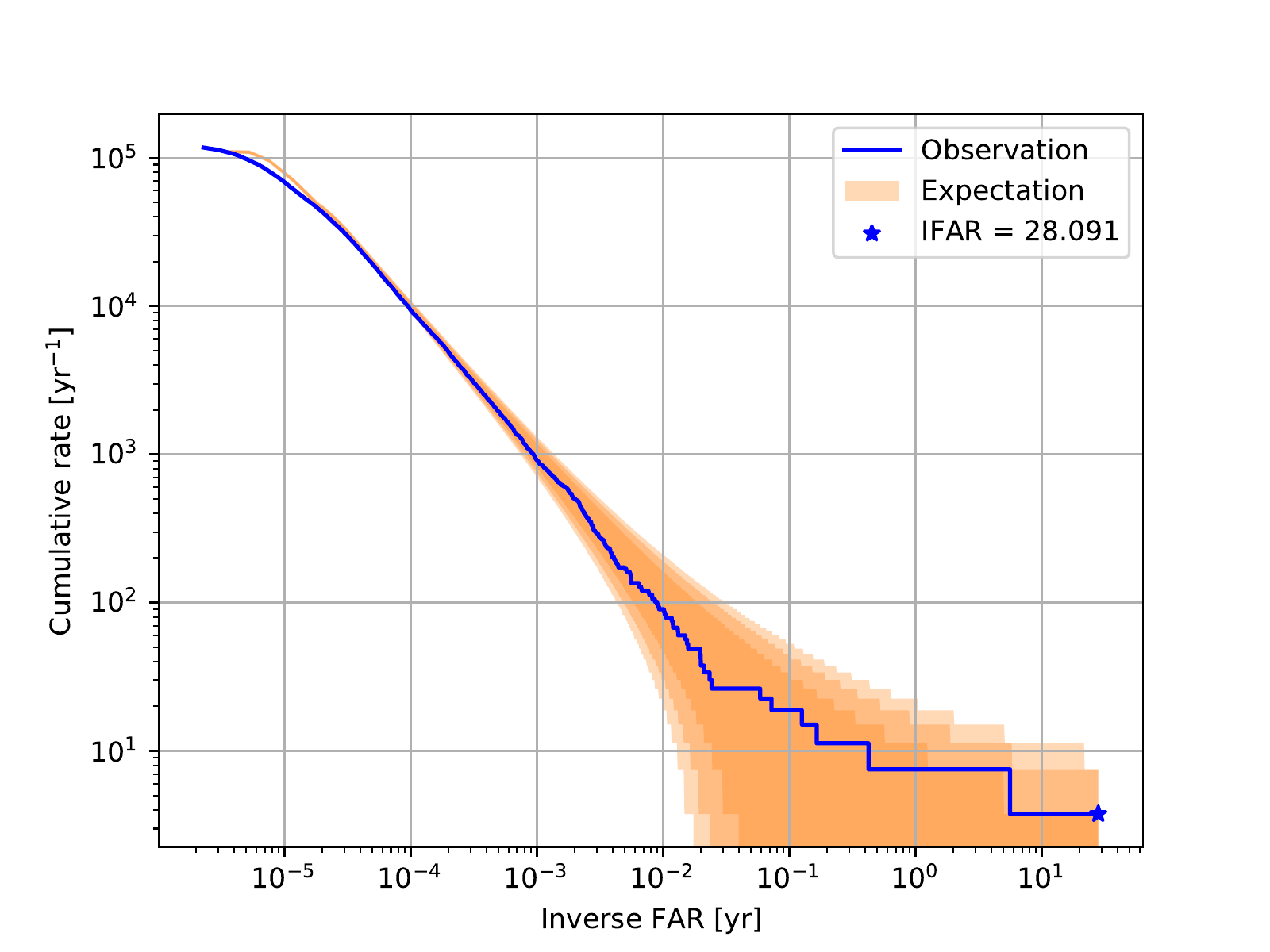}
    \caption{Modeling the turnover of the curve of the cumulative rate as a function of the IFAR in configuration $n^{\circ}$1 when we separate the associations by GBM spectral hardness and duration value.}
    \label{fig:model_rate}
\end{figure}

To check the validity of this modeling, one can compute the significance of the foreground in configuration $n^{\circ}$1 when we separate the analysis by GBM spectral value and duration. The results are represented in Figure~\ref{fig:model_rate}. To improve this modeling, one could define bins for each (spectral hardness-duration) pair (and not only duration as presented here) and apply a weight for each of these bins. This could be something to investigate for future analysis.

\bibliography{references}{}
\bibliographystyle{aasjournal}

\end{document}